\newif\ifpreprint%
\preprinttrue
\newcommand{\commonoptslist}%
	{%
	aps,%
	prd,%
	showpacs,%
	floatfix,%
	amsmath,%
	amssymb,%
	amsfonts,%
	nofootinbib,%
	superscriptaddress,%
	}%
\ifpreprint
\documentclass%
	[%
	preprint,%
	\commonoptslist%
	]{revtex4}
\else
\documentclass%
	[%
	twocolumn,%
	\commonoptslist{}%
	]{revtex4}
\fi
		\usepackage{ifthen}
		\usepackage{mathrsfs}		% for Script Letters 
		\usepackage[bbgreekl]{mathbbol}	% for double struck GREEK letters
		\usepackage{bbm}			% for double struck letters
		\usepackage{natbib}		% for references
		\usepackage[usenames]{color}	% for color
		\usepackage{graphicx}		% for displaying pictures
		\usepackage{epstopdf}
		\usepackage{subfig}		% for figures within figures
		\usepackage{abbrvate}		% for abbreviations
		\newcommand{\imgpath}{Images}
			\newabbr{AMSB}
			    {Anomaly Mediated Supersymmetry Breaking}
			\newabbr{EWSB}{electro-weak symmetry breaking}
			\newabbr{mAMSB}
			    {Minimal Anomaly Mediated Supersymmetry Breaking}
			\newabbr{GMSB}{Gauge Mediated Supersymmetry Breaking}
			\newabbr{mGMSB}
			    {Minimal Gauge Mediated Supersymmetry Breaking}
			\newabbr{LSP}{Lightest Supersymmetric Particle}
			\newabbr{LHC}{Large Hadron Collider}
			\newabbr{NLSP}
			    {Next-to Lightest Supersymmetric Particle}
			\newabbr{mSUGRA}{Minimal Supergravity}
			\newabbr{MSSM}{Minimal Supersymmetric Standard Model}
			\newabbr{NMSSM}
			    {Next-to Minimal Supersymmetric Standard Model}
			\newabbr{RGE}{Renormalization Group Equation}
			\newabbr{SM}{Standard Model}
			\newabbr{SUSY}{supersymmetry}
			\newabbr{SUSYLR}{supersymmetric left-right}
			\newabbr{UV}{ultra violet}
			\newabbr{VEV}{vacuum expectation value}
			\newcommand{\thismodel}{LR-AMSB}
			\newcommand{\eat}[1]{}
			\newcommand{\WSUSYLR}{W_{\text{SUSYLR}}}
			\newcommand{\DeltaBar}{\bar{\Delta}}
			\newcommand{\DeltaC}{\Delta^c}
			\newcommand{\DeltaBarC}{\bar{\Delta}^c}
			\newboolean{ulinescalar}
			\newcommand{\scalarunderline}{%
				\setboolean{ulinescalar}{true}%
				}%
			\newcommand{\noscalarunderline}{%
				\setboolean{ulinescalar}{false}%
				}%
			\newcommand{\scalar}[1]{%
				\ifthenelse{ \boolean{ulinescalar} }
					{
					\ifthenelse{ \equal{#1}{\DeltaC} }%
						{%
						\underline{\Delta}^c%
						}%
					{\ifthenelse{ \equal{#1}{\DeltaBarC}}%
						{%
						\underline{\bar{\Delta}}^c%
						}%
					{\ifthenelse{ \equal{#1}{\Phi}}%
						{%
						\underline{\Phi}%
						}%
						{%
						\underline{#1}%
						}}}%
					}
					{
					#1%
					}
				}%
				\newcommand{\DBarCpp}{\bar{\Delta}^{c ++}}

				\newcommand{\DCmm}{\Delta^{c --}}

			\newcommand{\Slight}{N}
			\newcommand{\Slightvev}{n}
			\newcommand{\Sheavy}{S}
			\newcommand{\kappaLight}{\kappa_\Slight}
			\newcommand{\kappaHeavy}{\kappa_\Sheavy}
			\newcommand{\lambdaLight}{\lambda_\Slight}
			\newcommand{\lambdaHeavy}{\lambda_\Sheavy}
				\newcommand{\I}{\mathbbm{i}}
				\newcommand{\EE}{\mathbbm{e}}
				\newcommand{\MP}{M_{\text{Pl}}}
				\newcommand{\MNP}{M_X}
				\newcommand{\Fphi}{F_{\phi}}
				\newcommand{\man}{m_{\text{an}}}
				\newcommand{\MSUSY}{M_\text{SUSY}}
				\newcommand{\Mmess}{M_\text{mess}}
				\newcommand{\MZ}{M_Z}
				\newcommand{\MW}{M_W}

			\newcommand{\pderiv}[3][]{%
				\frac{ \partial^{#1} #2}{ \partial {#3}^{#1} }%
				}
			\newcommand{\deriv}[3][]{%
				\frac{ d^{#1} #2}{ d {#3}^{#1} }%
				}
			\newcommand{\intOp}[2][]{\int \! d^{#1}#2 \;}
			\newcommand{\abs}[1]{ \mathopen{}\left| {#1}\right| }

			\DeclareMathOperator{\sgn}{sgn}
			\newcommand{\half}{\frac{1}{2}}			% 1/2
			\newcommand{\third}{\frac{1}{3}}			% 1/3
			\newcommand{\fourth}{\frac{1}{4}}		% 1/4
			\newcommand{\eighth}{\frac{1}{8}}		% 1/8
			\DeclareMathOperator{\Tr}{Tr}
			\newcommand{\inp}[2][0cm]{%
				\mathopen{}\left(%
					#2\parbox[h][#1]{0cm}{}\right)%
				}
			\newcommand{\inb}[2][0cm]{%
				\mathopen{}\left[%
					#2\parbox[h][#1]{0cm}{}\right]%
				}
			\newcommand{\inap}[2][0cm]{%
				\mathopen{}\left<%
					#2\parbox[h][#1]{0cm}{}\right>%
				}
			\newcommand{\nop}[1]{\mathopen{}\left.#1\right.%
				}
			\newcommand{\pfrac}[2]{%
				\mathopen{}\left(\frac{#1}{#2}\right)%
				}
				\newcommand{\Mu}{\text{M}}
		\newcommand{\E}[1]{\times 10^{#1}}
		\newcommand{\vev}[1]{\inap{#1}}
		\newcommand{\GSPNR}{GSPNR}
		\newcommand{\GSVNR}{GSVNR}
		\newcommand{\eq}[1]{Eq.~\eqref{Eq:#1}}
		\newcommand{\eqn}[1]{\eqref{Eq:#1}}
		\newcommand{\fig}[1]{Figure~\ref{Fig:#1}}
		\newcommand{\Sec}[1]{Section~\ref{Sec:#1}}
		\newcommand{\tbl}[1]{Table~\ref{Table:#1}}
		\newcommand{\app}[1]{Appendix~\ref{App:#1}}
		\newcommand{\bibpath}{../bibtex}
\begin{document}
%|%%%%%%%%%%%%%%%%%%%%%%%%%%%%%%%DO NOT EXCEED%%%%%%%%%%%%%%%%%%%%%%%%%%%%%%%|%

%%%%%%%%%%%%%%%%%%%%%%%%%%%%%%%%%%%%%%%%%%%%%%%%%%%%%%%%%%%%%%%%%%%%%%%%%%%%%%%
% * * * * * * * * * * * * * * * * * * * * * * * * * * * * * * * * * * * * * * *
\begin{abstract}
Superconformal anomalies provide an elegant and economical way to understand 
the soft breaking parameters in %\abbr[>-1+ic]{SUSY} 
SUSY models; however, 
implementing them leads to the several undesirable features including:
tachyonic sleptons and electroweak symmetry breaking problems in both the 
%\abbr{MSSM} 
MSSM and the %\abbr{NMSSM}
NMSSM.  Since these two theories also have the 
additonal problem of massless neutrinos, we have reconsidered the %\abbr{AMSB}
AMSB 
problems in a class of models that extends %\abbr{NMSSM} 
the NMSSM to explain small 
neutrino masses via the seesaw mechanism.  In a recent paper, we showed that 
for a class of minimal left-right extensions, a built-in mechanism 
exists which naturally solves the tachyonic slepton problem and provides new 
alternatives to the %\abbr{MSSM}
MSSM that also have automatic $R$-parity 
conservation. In this paper, we discuss how electroweak symmetry breaking arises 
in this model through an %\abbr{NMSSM}
NMSSM-like low energy theory with a singlet 
%\abbr{VEV}
VEV, induced by the structure of the left-right extension and of the 
right magnitude. We then study the phenomenological issues and find: the 
%\abbr{LSP}
LSP is an Higgsino-wino mix, new phenomenology for chargino decays to 
the %\abbr{LSP}
LSP, degenerate same generation sleptons and a potential for a mild 
squark-slepton degeneracy.  We also discuss possible collider signatures and 
the feasibility of dark matter in this model.
\end{abstract}
% * * * * * * * * * * * * * * * * * * * * * * * * * * * * * * * * * * * * * * *
%%%%%%%%%%%%%%%%%%%%%%%%%%%%%%%%%%%%%%%%%%%%%%%%%%%%%%%%%%%%%%%%%%%%%%%%%%%%%%%

%|%%%%%%%%%%%%%%%%%%%%%%%%%%%%%%%DO NOT EXCEED%%%%%%%%%%%%%%%%%%%%%%%%%%%%%%%|%
% <docproperties>
	\title{	Seesaw Extended %\abbr{MSSM} 
		MSSM
		and Anomaly Mediation without 
		Tachyonic Sleptons
		}
	\author{R. N. Mohapatra, N. Setzer, S. Spinner}
	\affiliation{%
		Department of Physics,
		Maryland Center for Fundamental Physics,
		University of Maryland,
		College Park, MD 20742,
		USA%
		}
	\date{January, 2008}
	\preprint{\vbox{ \hbox{UMD-PP-08-001} }}
	\pagestyle{plain}	% Page numbers bottom-center
	\abbrstyle{plain}
% </docproperties>

\maketitle

%|%%%%%%%%%%%%%%%%%%%%%%%%%%%%%%%DO NOT EXCEED%%%%%%%%%%%%%%%%%%%%%%%%%%%%%%%|%

%%%%%%%%%%%%%%%%%%%%%%%%%%%%%%%%%%%%%%%%%%%%%%%%%%%%%%%%%%%%%%%%%%%%%%%%%%%%%%%
%%%%%%%%%%%%%%%%%%%%%%%%%%%%%%%%%%%%%%%%%%%%%%%%%%%%%%%%%%%%%%%%%%%%%%%%%%%%%%%
\section{Introduction}
%%%%%%%%%%%%%%%%%%%%%%%%%%%%%%%%%%%%%%%%%%%%%%%%%%%%%%%%%%%%%%%%%%%%%%%%%%%%%%%
%%%%%%%%%%%%%%%%%%%%%%%%%%%%%%%%%%%%%%%%%%%%%%%%%%%%%%%%%%%%%%%%%%%%%%%%%%%%%%%

One of the leading candidates for TeV scale physics is the supersymmetric
extension of the \abbr{SM}\cite{Martin:1997ns} since it resolves an
outstanding \abbr{SM} conceptual issue: the gauge hierarchy problem (or 
why
$\MZ \ll \MP$ is stable under radiative corrections). It also leads to gauge
coupling unification as well as a candidate for dark matter of the universe if
two additional assumptions are made: a grand desert until $M \sim 10^{16}$ GeV
for gauge unification, and exact $R$-parity for dark matter. In addition it has 
the potential to explain the origin of spontaneous breaking of electroweak
symmetry.  Of course, \abbr{SUSY} has to be a broken symmetry to
conform with observations because no superpartner particles have been observed
yet.  Understanding the nature and origin of this \abbr{SUSY} breaking is a 
major
challenge which has commanded a great deal of attention.  An attractive and
elegant mechanism is to use the superconformal anomaly\cite{Randall:1998uk,
Giudice:1998xp} to break supersymmetry in the manner that has been dubbed
\abbr{AMSB}. \abbr{AMSB} provides an 
ultra-violet insensitive way to determine the soft \abbr{SUSY} breaking 
parameters\cite{Dine:2007me,Boyda:2001nh} as they depend
only on the TeV scale gauge Yukawa couplings of the low energy theory.
Consequently, it considerably reduces the number of arbitrary parameters of the
\abbr{SUSY} breaking sector. It also provides a heavy gravitino which has a 
number of cosmological advantages.

%|%%%%%%%%%%%%%%%%%%%%%%%%%%%%%%%DO NOT EXCEED%%%%%%%%%%%%%%%%%%%%%%%%%%%%%%%|%

A major problem of \abbr{AMSB} is that when implemented in the \abbr{MSSM}, it 
leads to negative slepton mass-squares---an unacceptable scenario since it leads 
to the breakdown of electric charge (sometimes called the tachyonic slepton 
problem). Another stumbling block to realistic \abbr{AMSB} model building is 
\abbr{EWSB}: the explicit $\mu$ term in the \abbr{MSSM} gives a $B\mu$ that is too large, 
while extensions like the 
\abbr{NMSSM} fail to generate a $\mu$ term that is large enough.  A number of attempts 
have been made to extend the \abbr{MSSM} in order to cure these 
problems\cite{Chacko:1999am,Gherghetta:1999sw,
Katz:1999uw,Pomarol:1999ie,Arkani-Hamed:2000xj,Allanach:2000gu,Jack:2000cd,
Carena:2000ad,Ibe:2004tg,Okada:2002mv}, usually with a focus on the tachyonic 
slepton problem.

%|%%%%%%%%%%%%%%%%%%%%%%%%%%%%%%%DO NOT EXCEED%%%%%%%%%%%%%%%%%%%%%%%%%%%%%%%|%

Since in \abbr{AMSB} models the \abbr{SUSY} breaking profile is crucially 
dependent on the low energy
theory, an interesting question arises as to whether \abbr{AMSB} still has the 
same
problems when the \abbr{MSSM} extended to accomodate neutrino masses. In a 
recent
paper\cite{Mohapatra:2007xq}, we pointed out that when the \abbr{MSSM} is 
minimally
extended to the \abbr{SUSYLR} model with $B - L = 2$
triplets to implement the seesaw mechanism, the low energy particle content
and interaction profile changes just enough to cure the negative slepton mass
square problem. A key feature responsible for this cure is the appearance of
a naturally light $SU(2)_L$ triplet and a doubly-charged singlet which have
leptonic Yukawa interactions. In Ref.\cite{Mohapatra:2007xq}, we
explained how \abbr{SUSYLR} fixes the tachyonic slepton problem of \abbr{AMSB} 
and also noted some of the gross distinguishing features of the model---such 
as the appearance of 
$B - L = 2$ triplets, doubly-charged Higgs bosons,
and a pair of additional heavy Higgs doublets all with masses around the mass 
scale of conformal \abbr{SUSY} breaking, $F_\phi$---typically in the tens of TeVs.  
Since then another paper has explored the 
relationship of neutrinos and \abbr{AMSB} in the context of defltected 
\abbr{AMSB} \cite{Mohapatra:2007js}.

%|%%%%%%%%%%%%%%%%%%%%%%%%%%%%%%%DO NOT EXCEED%%%%%%%%%%%%%%%%%%%%%%%%%%%%%%%|%

In this paper, which should be viewed as a sequel to
ref.\cite{Mohapatra:2007xq}, we attempt to present a complete 
phenomenologically acceptable model addressing questions such as \abbr{EWSB}, 
 and dark matter. A summary of our results is as 
follows:
\begin{itemize}
\item{We show that the model below the $F_\phi$ scale is the \abbr{NMSSM} with 
a singlet superpotential mass term, $\mu_N$.  This term is necessary for 
\abbr{EWSB} and can arise from the \abbr{SUSYLR} framework necessary for the solution 
to the tachyonic slepton problem.}
\item{One implication of the similarity to the \abbr{NMSSM} below the TeV scale is
that the magnitude of the $B\mu$-term is of the desired magnitude.}
\item{We present the sparticle spectrum of the model for a generic
choice of the parameters and in particular we display the lightest
superparticle which can be the dark matter of the universe.  We find that 
same generation sleptons are degenerate 
and that a possibility exists for degenerate sleptons and squarks.}
\item{We find that the mass difference between the chargino and the 
lightest neutralino in our model is much larger 
than the \abbr{mAMSB} models where a universal scalar mass corrects 
the tachyonic slepton mass problem.}
\end{itemize}

%|%%%%%%%%%%%%%%%%%%%%%%%%%%%%%%%DO NOT EXCEED%%%%%%%%%%%%%%%%%%%%%%%%%%%%%%%|%

The paper is organized as follows: in \Sec{SUSYLR+AMSB.Review}, we review the
basic ingredients of the \abbr{SUSYLR} model that is the framework of our 
discussion;
in \Sec{TeV.Model.Spectrum}, we show the multi-TeV scale spectrum of the 
model and
discuss how it solves the negative slepton mass square problem of the model;
in \Sec{Low.Scale.Theory}, we discuss the effective theory below the $F_\phi$ TeV
scale and show how
electroweak symmetry breaking arises. In \Sec{Sparticle.Masses}, we display
the sparticle spectrum and compare it with that in some other benchmark 
\abbr{SUSY}
models with different \abbr{SUSY} breaking mechanisms. For the allowed 
parameter space of our model, we find a Higgsino-wino mixture to be the 
\abbr{LSP} and mention its prospects as the dark matter of the universe.  
We finish with a brief discussion of the ultraviolet consequences of this 
model in \Sec{Miscelleneous} and a conclusion.

%|%%%%%%%%%%%%%%%%%%%%%%%%%%%%%%%DO NOT EXCEED%%%%%%%%%%%%%%%%%%%%%%%%%%%%%%%|%

%%%%%%%%%%%%%%%%%%%%%%%%%%%%%%%%%%%%%%%%%%%%%%%%%%%%%%%%%%%%%%%%%%%%%%%%%%%%%%%
%%%%%%%%%%%%%%%%%%%%%%%%%%%%%%%%%%%%%%%%%%%%%%%%%%%%%%%%%%%%%%%%%%%%%%%%%%%%%%%
\section{Minimal \abbr{SUSYLR} Model Cures the problems of \abbr{AMSB}: a brief
review}
\label{Sec:SUSYLR+AMSB.Review}
%%%%%%%%%%%%%%%%%%%%%%%%%%%%%%%%%%%%%%%%%%%%%%%%%%%%%%%%%%%%%%%%%%%%%%%%%%%%%%%
%%%%%%%%%%%%%%%%%%%%%%%%%%%%%%%%%%%%%%%%%%%%%%%%%%%%%%%%%%%%%%%%%%%%%%%%%%%%%%%

\scalarunderline

%|%%%%%%%%%%%%%%%%%%%%%%%%%%%%%%%DO NOT EXCEED%%%%%%%%%%%%%%%%%%%%%%%%%%%%%%%|%

In generic \abbr{AMSB} models the soft \abbr{SUSY} breaking parameters 
associated 
with the superfield combination $\Phi_i \Phi^{j*}$ are determined by the 
anomalous dimensions $\gamma^i_j \inp{g_a, Y^{\ell m n}}$ and the scaling 
functions $\beta_g^a \inp{g_b,Y^{ijk}}$, $\beta_Y^{ijk} \inp{g_a,Y^{\ell m n}}$ 
of the low energy theory:
\begin{align}
\label{Eq:AMSB.scalar.mass}
\inp{m^2}^i_j
	& =	- \fourth \abs{\Fphi}^2
		\inb{
			  \half \pderiv{\gamma^i_j}{g_a} \beta_{g}^a 
			+ \pderiv{\gamma^i_j}{Y^{\ell m n}} \beta_Y^{\ell m n}
			+ \text{h.c.}
		}
	\\
\label{Eq:AMSB.trilinear.A}
a^{ijk}	& = \beta_{Y}^{ijk} \Fphi
	\\
\label{Eq:AMSB.gaugino.mass}
M_{\lambda_a}
	& = \frac{\beta_{g}^a}{g_a} \Fphi
\end{align}
Here $\Fphi$ is the \abbr{SUSY} breaking scale in the gauge where the 
conformal compensator $\phi$ has the form
\begin{equation}
\phi = 1 + \Fphi \theta^2
\end{equation}
%
%|%%%%%%%%%%%%%%%%%%%%%%%%%%%%%%%DO NOT EXCEED%%%%%%%%%%%%%%%%%%%%%%%%%%%%%%%|%
%
with $\Fphi$ as an input parameter having a value in the $10$s of TeV range. 
The remainder of our notational conventions can be found in
\app{Notation.Conventions}.

%|%%%%%%%%%%%%%%%%%%%%%%%%%%%%%%%DO NOT EXCEED%%%%%%%%%%%%%%%%%%%%%%%%%%%%%%%|%

It is clear from \eq{AMSB.scalar.mass} that when this formula is applied to
the \abbr{MSSM}, the slepton mass-squares are negative due to the positive
(asymptotically non-free) $SU(2) \times U(1)_Y$ gauge couplings' $\beta$
functions and the nearly zero lepton Yukawa couplings\footnote{While the
Yukawa coupling of $\tau$ might be significant, the first and second
generation leptons have negligible Yukawa couplings}.  As pointed out in
Ref.~\cite{Mohapatra:2007xq}, this problem is cured by extending the \abbr{MSSM} to 
\abbr{SUSYLR} due
to the following property: the effective theory below the seesaw scale $v_R$
contains a set of $SU(2)_L$ triplets and doubly-charged fields, both 
having Yukawa couplings to the left- and right-handed leptons 
respectively. Their masses are naturally in the multi-TeV range 
despite the high 
seesaw scale due to an accidental global symmetry of the 
theory\cite{Aulakh:1998nn,Chacko:1997cm}. Furthermore, provided these new 
couplings are of order $1$, the slepton masses
squares can be made positive.  Thus, \abbr{SUSYLR} not only explains the small
neutrino masses by means of the seesaw mechanism, but its marriage with 
\abbr{AMSB}
cures the negative slepton mass-square problem.  The resulting theory combines
the predictive power of \abbr{AMSB}, explains neutrino masses, and retains a 
natural
dark matter candidate due to the theory's automatic conservation of $R$-Parity 
below the right-handed scale. It also contains a mechanism for generating an 
appropriate singlet \abbr{VEV} in the effective low energy \abbr{NMSSM}-like 
superpotential.  In the following subsections, we fill in the details.

%|%%%%%%%%%%%%%%%%%%%%%%%%%%%%%%%DO NOT EXCEED%%%%%%%%%%%%%%%%%%%%%%%%%%%%%%%|%

%%%%%%%%%%%%%%%%%%%%%%%%%%%%%%%%%%%%%%%%%%%%%%%%%%%%%%%%%%%%%%%%%%%%%%%%%%%%%%%
% % % % % % % % % % % % % % % % % % % % % % % % % % % % % % % % % % % % % % % %
\subsection{The Left-Right Model}
\label{Sec:Left-Right}
% % % % % % % % % % % % % % % % % % % % % % % % % % % % % % % % % % % % % % % %
%%%%%%%%%%%%%%%%%%%%%%%%%%%%%%%%%%%%%%%%%%%%%%%%%%%%%%%%%%%%%%%%%%%%%%%%%%%%%%%

The particle content of a \abbr{SUSYLR} model is shown in \tbl{QM Numbers}. 
As the
model is left-right symmetric, it contains both left- and right-handed higgs
bosons---in this case $B - L = \pm 2$ triplets so that $R$-parity may be
preserved (a task for which $B - L = 1$ doublets are not suitable).  The
presence of both $SU(2)_L$ and $SU(2)_R$ triplets means that parity is a good
symmetry until $SU(2)_R$ breaks.  While the seesaw mechanism may be achieved
with only $SU(2)_R$ higgs fields, demanding parity forces the presence of
left-handed triplets.  The inclusion of both these fields then leads to
positive left- and right-handed slepton masses.

%|%%%%%%%%%%%%%%%%%%%%%%%%%%%%%%%DO NOT EXCEED%%%%%%%%%%%%%%%%%%%%%%%%%%%%%%%|%
	
\begin{table}[ht]
\begin{center}
\begin{tabular}{|c|c|}
\hline\hline
Fields		& $SU(3)^c \times SU(2)_L \times SU(2)_R \times U(1)_{B-L}$
	\\
\hline
$Q$		& $(3   ,   2   ,   1   ,   +\third )$
	\\
$Q^c$		& $(\bar{3},1   ,   2   ,   -\third )$
	\\
$L$		& $(1   ,   2   ,   1   ,   -1  )$
	\\
$L^c$		& $(1   ,   1   ,   2   ,   +1  )$
	\\
$\Phi_a$		& $(1   ,   2   ,   2   ,   0   )$
	\\
$\Delta$		& $(1   ,   3   ,   1   ,   +2  )$
	\\
$\DeltaBar$	& $(1   ,   3   ,   1   ,   -2  )$
	\\
$\DeltaC$  	& $(1   ,   1   ,   3   ,   -2  )$
	\\
$\DeltaBarC$	& $(1   ,   1   ,   3   ,   +2  )$
	\\
$\Sheavy,\Slight$
		& $(1   ,   1   ,   1   ,   0   )$
	\\
\hline\hline
\end{tabular}
\end{center}
%|%%%%%%%%%%%%%%%%%%%%%%%%%%%%%%%DO NOT EXCEED%%%%%%%%%%%%%%%%%%%%%%%%%%%%%%%|%
\caption{%
Assignment of the matter and Higgs fields' representations of the
left-right symmetry group (except for $U(1)_{B-L}$ where the charge under
that group is given.)%
}
\label{Table:QM Numbers}
\end{table}
%|%%%%%%%%%%%%%%%%%%%%%%%%%%%%%%%DO NOT EXCEED%%%%%%%%%%%%%%%%%%%%%%%%%%%%%%%|%

To be explicit, the fields of \tbl{QM Numbers} transform under parity as
\begin{align*}
Q		& \leftrightarrow - \I \tau_2 \nop{Q^c}^*	&
L		& \leftrightarrow - \I \tau_2 \nop{L^c}^*	&
\Phi_a		& \rightarrow \Phi_a^\dagger			\\
\Delta		& \leftrightarrow \nop{\DeltaC}^\dagger		&
\DeltaBar	& \leftrightarrow \nop{\DeltaBarC}^\dagger	&
\Sheavy,\Slight
		& \rightarrow \Sheavy^*,\Slight^*
\end{align*}
so that the fully parity symmetric superpotential is
\begin{align}
\label{Eq:SuperW.SUSYLR}
\WSUSYLR
	& = W_{\text{Y}} + W_{\text{H}} + W_{\text{\GSPNR}} + W_{\text{\GSVNR}}
\end{align}
%
%|%%%%%%%%%%%%%%%%%%%%%%%%%%%%%%%DO NOT EXCEED%%%%%%%%%%%%%%%%%%%%%%%%%%%%%%%|%
with
\begin{align}
W_{\text{Y}}
	& =	   \I y_{Q}^a Q^T \tau_2 \Phi_a Q^c
		+ \I y_{L}^a L^T \tau_2 \Phi_a L^c
		+ \I f_c L^{cT} \tau_2 \DeltaC L^c
		+ \I f L^T \tau_2 \Delta L
\label{Eq:SuperW.SUSYLR.Yuk}
	\\
W_{\text{H}}
	& =	\inp{
			M_{\Delta} \phi - \lambdaHeavy \Sheavy
		   } \inb{
			  \Tr\inp{\DeltaC \DeltaBarC}
			+ \Tr\inp{\Delta \DeltaBar}
			}
		+ M_{\Sheavy}^2 \phi^2 \Sheavy
		+ \half \mu_{\Sheavy} \phi \Sheavy^2
		+ \third \kappaHeavy \Sheavy^3
%|%%%%%%%%%%%%%%%%%%%%%%%%%%%%%%%DO NOT EXCEED%%%%%%%%%%%%%%%%%%%%%%%%%%%%%%%|%
\notag	\\
	& \quad {}
		+ \lambdaLight^{ab} \Slight
			\Tr\inp{\Phi_a^T \tau_2 \Phi_b \tau_2}
	+ \third \kappaLight \Slight^3
\label{Eq:SuperW.SUSYLR.Higgs}
\end{align}
\begin{align}
W_{\text{\GSPNR}}
	& =	  \frac{\lambda_A}{\MNP \phi} \Tr^2\inp{\Delta \DeltaBar}
		+ \frac{\lambda_A^c}{\MNP \phi} \Tr^2\inp{\DeltaC \DeltaBarC}
\notag	\\
	& \quad {}
		+ \frac{\lambda_B}{\MNP \phi}
			\Tr\inp{\Delta \Delta} \Tr\inp{\DeltaBar \DeltaBar}
		+ \frac{\lambda_B^c}{\MNP \phi}
			\Tr\inp{\DeltaC \DeltaC} \Tr\inp{\DeltaBarC \DeltaBarC}
%|%%%%%%%%%%%%%%%%%%%%%%%%%%%%%%%DO NOT EXCEED%%%%%%%%%%%%%%%%%%%%%%%%%%%%%%%|%
\notag	\\
	& \quad {}
		+ \frac{\lambda_C}{\MNP \phi}
			\Tr\inp{\Delta \DeltaBar} \Tr\inp{\DeltaC \DeltaBarC}
\notag	\\
	& \quad {}
		+ \frac{ \lambda_{\mathbf{\Sheavy}} }{ \MNP \phi }
			\Tr\inp{\Delta \DeltaBar} \Sheavy^2
		+ \frac{ \lambda_{\mathbf{\Sheavy}}^c }{ \MNP \phi }
			\Tr\inp{\DeltaC \DeltaBarC} \Sheavy^2
		+ \cdots
\label{Eq:SuperW.SUSYLR.nr.gsp}
	\\
W_{\text{\GSVNR}}
	& =
		  \frac{\lambda_D}{\MP \phi}
			\Tr\inp{\Delta \Delta}
			\Tr\inp{\DeltaC \DeltaC}
		+ \frac{\bar\lambda_D}{\MP \phi}
			\Tr\inp{\DeltaBar \DeltaBar}
			\Tr\inp{\DeltaBarC \DeltaBarC}
%|%%%%%%%%%%%%%%%%%%%%%%%%%%%%%%%DO NOT EXCEED%%%%%%%%%%%%%%%%%%%%%%%%%%%%%%%|%
\notag	\\
	& \quad {}
		+ \frac{\inp{\lambda_\sigma}^{ab}}{\MP \phi}
			\Tr\inp{\Delta \DeltaBar}
			\Tr\inp{ \Phi_a^T \tau_2 \Phi_b \tau_2 }
		+ \frac{\inp{\lambda_\sigma^c}^{ab}}{\MP \phi}
			\Tr\inp{\DeltaC \DeltaBarC}
			\Tr\inp{ \Phi_a^T \tau_2 \Phi_b \tau_2 }
%|%%%%%%%%%%%%%%%%%%%%%%%%%%%%%%%DO NOT EXCEED%%%%%%%%%%%%%%%%%%%%%%%%%%%%%%%|%
\notag	\\
	& \quad {}
		+ \frac{2\lambda_\alpha \epsilon^{ab}}{\MP \phi}
		   \Tr\inp{\Delta \Phi_a \tau_2 \Phi_b^T \tau_2 \DeltaBar}
		+ \frac{2\lambda_\alpha^c \epsilon^{ab}}{\MP \phi}
		   \Tr\inp{\DeltaC \tau_2 \Phi_a^T \tau_2 \Phi_b \DeltaBarC}
\notag	\\
	& \quad {}
		+ \frac{ \lambda_{\mathbf{\Slight}} }{ \MP \phi }
			\Tr\inp{\Delta \DeltaBar} \Slight^2
		+ \frac{ \lambda_{\mathbf{\Slight}}^c }{ \MP \phi }
			\Tr\inp{\DeltaC \DeltaBarC} \Slight^2
%|%%%%%%%%%%%%%%%%%%%%%%%%%%%%%%%DO NOT EXCEED%%%%%%%%%%%%%%%%%%%%%%%%%%%%%%%|%
\notag	\\
	& \quad {}
		+ \frac	{ \lambda_{\mathbf{\MakeLowercase{\Sheavy}}} }
			{ \MP \phi }
			\Tr\inp{\Phi_a^T \tau_2 \Phi_b \tau_2} \Sheavy^2
		+ \frac{ \lambda_\Mu }{ \MP \phi } \Sheavy^2 \Slight^2
		+ \cdots
\label{Eq:SuperW.SUSYLR.nr.gsv}
\end{align}
Furthermore, parity demands that the couplings be related as
\begin{align*}
y_Q^a		& = \inp{y_Q^a}^\dagger		&
y_L^a		& = \inp{y_L^a}^\dagger		&
f		& = f_c^*			&
M_\Delta		& = M_\Delta^*			\\
\lambdaHeavy	& = \lambdaHeavy^*		&
M_{\Sheavy}^2	& = \inp{M_{\Sheavy}^2}^*	&
\mu_{\Sheavy}	& = \mu_{\Sheavy}^*		&
\kappaHeavy	& = \kappaHeavy^*		\\
		&				&
\lambdaLight	& = \lambdaLight^\dagger		&
\kappaLight	& = \kappaLight^*
\end{align*}
We have also imposed a discrete $\mathbbm{Z}_3$ symmetry on \eq{SuperW.SUSYLR}
with
%
%|%%%%%%%%%%%%%%%%%%%%%%%%%%%%%%%DO NOT EXCEED%%%%%%%%%%%%%%%%%%%%%%%%%%%%%%%|%
%
\begin{equation}
\begin{aligned}
(Q,Q^c,L,L^c,\Delta,\DeltaC,\Phi_a,\Slight)
	& \rightarrow \EE^{2 \I \pi /3}
		(Q,Q^c,L,L^c,\Delta,\DeltaC,\Phi_a,\Slight),
	\\
(\DeltaBar,\DeltaBarC)
	& \rightarrow \EE^{4 \I \pi /3} (\DeltaBar,\DeltaBarC)
\end{aligned}
\label{Eq:SUSYLR.Fields.Z3.Transformation}
\end{equation}
%
%|%%%%%%%%%%%%%%%%%%%%%%%%%%%%%%%DO NOT EXCEED%%%%%%%%%%%%%%%%%%%%%%%%%%%%%%%|%
%
and $\Sheavy$ invariant.  This symmetry is necessary to keep one singlet
light below the right-handed scale since it forbids terms such as
\begin{equation}
W_{\not{\mathbbm{Z}}_3}
	=	  \kappa_{12} \Sheavy \Slight^2
		+ \kappa_{21} \Sheavy^2 \Slight
		+ \lambda^c_{\Slight} \Slight \Tr \inp{\DeltaC \DeltaBarC}
% the names come from imagining \Sheavy and \Slight form a vector so that
% you could write their interactions as
% (\Sheavy, \Slight) \kappa \diag{\Sheavy, \Slight} (\Sheavy, \Slight)^T
% where \kappa is a 2x2 matrix
\label{Eq:Super.Potential.Z3.violating}
\end{equation}
which would generate a large, $\mathcal{O}(v_R)$, \abbr[>-1+ic]{SUSY} mass for 
$\Slight$.  Yet because it is a
global symmetry, it will be violated by gravitational
effects\footnote{For example, if a particle charged under this symmetry falls
into a blackhole, there is no way to ascertain the amount of this charge the
blackhole contains.  This can be contrasted with a gauged symmetry where
Gauss's law may be utilized to determine the charge enclosed} leading to
\eq{SuperW.SUSYLR} containing the non-renormalizable terms of 
\eq{SuperW.SUSYLR.nr.gsv} (which are accordingly suppressed by the
planck scale $\MP$).

The superpotential \eq{SuperW.SUSYLR} must also contain the additional
non-renormalizable terms given by \eq{SuperW.SUSYLR.nr.gsp} if the theory
is to preserve $R$-parity and be phenomenologically
viable\cite{Aulakh:1998nn,Chacko:1997cm}.  These terms preserve the
$\mathbbm{Z}_3$ symmetry and are therefore suppressed by the next
new scale of physics, which we have chosen to call $\MNP$.  We will show that
it is possible to fix $\MNP$ in \Sec{TeV.Model.Spectrum}, where we consider
the $\Fphi$ scale theory.

Meanwhile, the Higgs potential given by \eq{SuperW.SUSYLR.Higgs} dictates
that the \abbr{VEV} for the right-handed superfields are
%
%|%%%%%%%%%%%%%%%%%%%%%%%%%%%%%%%DO NOT EXCEED%%%%%%%%%%%%%%%%%%%%%%%%%%%%%%%|%
%
\begin{align}
\label{Eq:vev.S.heavy}
\vev{\Sheavy}
	& =	\frac{ M_{\Delta} }{ \lambdaHeavy } \phi
	\\
\label{Eq:vev.Dc.Dbarc}
\vev{\DeltaC} \vev{\DeltaBarC}
	& = \vev{\Sheavy}
		\inp{
			  \frac{M_{\Delta} \kappaHeavy }{ \lambdaHeavy^2 }
			+ \frac{ \mu_{\Sheavy} }{ \lambdaHeavy }
		} \phi + \frac{M_\Sheavy^2}{\lambdaHeavy} \phi^2
\end{align}
With $M_\Delta \sim \mu_\Sheavy \sim v_R \sim 10^{11}$ GeV, where $v_R$ is the 
right-handed breaking scale.
\eq{vev.S.heavy} should be evident from the form of the superpotential;
\eq{vev.Dc.Dbarc} requires \eq{SuperW.SUSYLR.Higgs} to be recast as
%
%|%%%%%%%%%%%%%%%%%%%%%%%%%%%%%%%DO NOT EXCEED%%%%%%%%%%%%%%%%%%%%%%%%%%%%%%%|%
%
\begin{equation}
W_{\text{H}}
	\supset	\inb{
			- \lambdaHeavy  \Tr\inp{\DeltaC \DeltaBarC}
			+ M_\Sheavy^2 \phi^2
			+ \half \mu_{\Sheavy} \phi \Sheavy
			+ \third \kappaHeavy \Sheavy^2
		} \Sheavy
\end{equation}

The non-renormalizable terms will shift the right-handed scale \abbr{VEV}s
by at most $\sim M_{\Delta}^2 / \MNP \ll M_{\Delta}$ so they may be safely be
ignored.  The theory then remains \abbr{UV} insensitive below
$v_R$\cite{Pomarol:1999ie} and hence respects the \abbr{AMSB} trajectory 
below this
scale. Yet even though the particles remain on their \abbr{AMSB} trajectory, 
the
negative slepton mass-squares problem is still solved due to the additional
low-scale yukawa couplings $f$ and $f_c$.

%|%%%%%%%%%%%%%%%%%%%%%%%%%%%%%%%DO NOT EXCEED%%%%%%%%%%%%%%%%%%%%%%%%%%%%%%%|%

To see why these yukawas survive, consider the Higgs sector of
\eq{SuperW.SUSYLR} before $SU(2)_R$ breaks and setting the
non-renormalizable terms to zero---essentially leaving just the terms in
\eq{SuperW.SUSYLR.Higgs}.  This superpotential has a complexified $U(6)$
symmetry\footnote{A
complexified $U(6)$ is a $U(6)$ with its parameters taken
to be complex.  Its existence in \eq{SuperW.SUSYLR.Higgs} can be seen
by defining two new fields $\mathbb{\Delta} \equiv (\Delta,\DeltaC)$ and
$\bar{\mathbb{\Delta}} \equiv (\DeltaBar,\DeltaBarC)$---which are complex
$6$-vectors---and combining the trace over each separately to
$\Tr\inp{\mathbb{\Delta}\bar{\mathbb{\Delta}}}$}
involving the $\Delta$'s and the $\DeltaC$'s
(similar symmetry arguments are discussed in \cite{Dutta:1998bn}, but because
the authors used a parity odd singlet, there was only a complexified $U(3)$
symmetry).  When $SU(2)_R$ breaks, the $U(6)$ is reduced to a $U(5)$ yielding
$22$ real degrees of freedom that are massless.  The $D$-terms and the gauge
fields consume $6$ of these, leaving a total of $16$ massless modes.  The
surviving $16$ massless real degrees of freedom are the two doubly-charged
$SU(2)_L$ singlets and the two left-handed triplets.

%|%%%%%%%%%%%%%%%%%%%%%%%%%%%%%%%DO NOT EXCEED%%%%%%%%%%%%%%%%%%%%%%%%%%%%%%%|%

Only the non-renormalizable terms of Eqs.~\eqn{SuperW.SUSYLR.nr.gsp} and
\eqn{SuperW.SUSYLR.nr.gsv} break the $U(6)$ symmetry, and therefore the
mass of the Higgsino must be
\begin{equation}
\mu_{\Delta,\DeltaBar} \sim \mu_{DC} \sim \frac{v_R^2}{\MNP}
\end{equation}
The \abbr{SUSY} breaking bilinear terms
generated by \abbr{AMSB} will 
force these masses
to be at least $\Fphi$ giving
\begin{equation}
\MNP \lesssim \frac{v_R^2}{\Fphi}.
\label{Eq:new.physics.expression}
\end{equation}
Thus, the scale of new physics is determined by the right-handed scale
and the \abbr{SUSY} breaking scale.

%|%%%%%%%%%%%%%%%%%%%%%%%%%%%%%%%DO NOT EXCEED%%%%%%%%%%%%%%%%%%%%%%%%%%%%%%%|%

The mass matrix for the left-handed triplets and doubly-charged Higgses have a 
similar form, here we state the doubly-charged matrix:
\begin{equation}
\label{Eq:DC.Mass}
{\cal M}_{DC}
	= \mu^2_{DC} 
	  \begin{pmatrix}
	  	1			& 1-\epsilon_\Delta	\\
		1-\epsilon_\Delta 		& 1
	  \end{pmatrix}
\end{equation}
where $\mu_{DC} \simeq F_\phi$ and $\epsilon_\Delta 
=1 - \frac{B_\Delta}{\mu_{DC}}$.  The eigenvalues of this mass matrix are 
$m_{DC}^2 = \epsilon_\Delta \mu^2_{DC}$ and 
$M_{DC}^2 = 2 \mu^2_{DC}$. 
Since $\epsilon_\Delta$ depends on $\mu_{DC}$, and $\mu_{DC}$ can be adjusted
through the coupling it contains, one doubly-charged 
Higgs can be made light. On the whole, we 
expect the two doubly charged scalar masses to be above $1$ TeV (for the 
lighter one) and $F_\phi$ (for the heavier one). Note that there is no 
such splitting between the fermionic partners, which remain heavy with a mass of 
about $\mu_{DC}$.  A similar argument applies to the left-handed triplets.

%|%%%%%%%%%%%%%%%%%%%%%%%%%%%%%%%DO NOT EXCEED%%%%%%%%%%%%%%%%%%%%%%%%%%%%%%%|%

Finally, because the masses of the $SU(2)_L$ triplets and the doubly-charged
particles will be around $\Fphi$, they are of the correct size to influence
the low-scale theory: if the masses had been large,
$F_\phi \ll  \mu_{DC} \ll v_R$, then they would have merely introduced another
trajectory preserving threshold that decoupled from the low scale theory.
However, because these particles remain in the low-scale theory, the effect of
their couplings is important.  For the sleptons the relevant terms are
\begin{equation}
	W \supset f_c \DCmm e^c e^c + \I f L^T \tau_2 \Delta L
\end{equation}
%
%|%%%%%%%%%%%%%%%%%%%%%%%%%%%%%%%DO NOT EXCEED%%%%%%%%%%%%%%%%%%%%%%%%%%%%%%%|%
%
with the surviving yukawa couplings $f_c$ and $f$ providing positive
mass-squares to the scalar leptons\footnote{Note that slepton mass
squares can also be positive for theories with a right handed scale lower than
$10^{11}$ GeV. We choose the high scale version since neutrino masses in this
case do not require any fine tuning of Yukawa couplings.}

%|%%%%%%%%%%%%%%%%%%%%%%%%%%%%%%%DO NOT EXCEED%%%%%%%%%%%%%%%%%%%%%%%%%%%%%%%|%

To make this explicit we write down the slepton masses with the contributions
of these additional interactions (taking the
$SU(2)_L \times U(1)_{Y}$ gauge couplings to be $g_2$ and $g_1$ respectively):
\begin{align}
m_{e^c}^2
	& = \half \frac{ \abs{F_\phi}^2 }{ \inp{16 \pi^2}^2 }
		\left[ \vphantom{\frac{ \abs{F_\phi}^2 }{ \inp{16 \pi^2}^2 }}
			  8 f_c^\dagger \inp{Y_L^a}^T \inp{Y_L^a}^* f_c
			+ 12 \inp{Y_L^a}^\dagger f f^\dagger Y_L^a
	\right.
\notag	\\
%|%%%%%%%%%%%%%%%%%%%%%%%%%%%%%%%DO NOT EXCEED%%%%%%%%%%%%%%%%%%%%%%%%%%%%%%%|%
	& \quad {}
		+ 8 f_c^\dagger f_c
			\inb{
				  \inp{Y_L^a}^\dagger Y_L^a
				+ 4 f_c^\dagger f_c
				+ \Tr\inp{f_c^\dagger f_c}
			}
		+ 4 \inp{Y_L^a}^\dagger Y_L^a
			\inb{
				  \inp{Y_L^b}^\dagger Y_L^b
				+ 2 f_c^\dagger f_c
			}
%|%%%%%%%%%%%%%%%%%%%%%%%%%%%%%%%DO NOT EXCEED%%%%%%%%%%%%%%%%%%%%%%%%%%%%%%%|%
\notag	\\
	& \quad {}
		+ 2 \inp{Y_L^a}^\dagger Y_L^b
			\inb{
				  2 \inp{Y_L^b}^\dagger Y_L^a
				+ \Tr\inp{
					  3 \inp{Y_Q^b}^\dagger Y_Q^a
					+ \inp{Y_L^b}^\dagger Y_L^a
					}
				+ 4 \inp{\lambdaLight^{cb}}^*
					\lambdaLight^{ca}
			}
%|%%%%%%%%%%%%%%%%%%%%%%%%%%%%%%%DO NOT EXCEED%%%%%%%%%%%%%%%%%%%%%%%%%%%%%%%|%
\notag	\\
	& \quad \left. {}
		- 2 g_1^2
			\inp{
				  24 f_c^\dagger f_c
				+ 3 \inp{Y_L^a}^\dagger Y_L^a
				+ 26 g_1^2
			}
		- 6 g_2^2 \inp{Y_L^a}^\dagger Y_L^a
		+ \text{h.c.}
		\vphantom{\frac{ \abs{F_\phi}^2 }{ \inp{16 \pi^2}^2 }}
	\right]
\label{Eq:Right.Selectron.Mass}
\end{align}
\begin{align}
%|%%%%%%%%%%%%%%%%%%%%%%%%%%%%%%%DO NOT EXCEED%%%%%%%%%%%%%%%%%%%%%%%%%%%%%%%|%
m_{L}^2
	& = \half \frac{ \abs{F_\phi}^2 }{ \inp{16 \pi^2}^2 }
		\left[ \vphantom{\frac{ \abs{F_\phi}^2 }{ \inp{16 \pi^2}^2 }}
			  6 f\inp{Y_L^a}^T \inp{Y_L^a}^* f^\dagger
			+ 4 Y_L^a f_c^\dagger f_c \inp{Y_L^a}^\dagger
		\right.
\notag	\\
%|%%%%%%%%%%%%%%%%%%%%%%%%%%%%%%%DO NOT EXCEED%%%%%%%%%%%%%%%%%%%%%%%%%%%%%%%|%
	& \quad {}
		+ 6 \inb{
			  \inp{Y_L^a}^\dagger Y_L^a
			+ 12 f f^\dagger
			+ 2 \Tr\inp{f^\dagger f}
			} f f^\dagger
		+ 2 \inb{
			  Y_L^b \inp{Y_L^b}^\dagger
			+ 3 f f^\dagger
			} Y_L^a \inp{Y_L^a}^\dagger
\notag	\\
%|%%%%%%%%%%%%%%%%%%%%%%%%%%%%%%%DO NOT EXCEED%%%%%%%%%%%%%%%%%%%%%%%%%%%%%%%|%
	& \quad {}
		+ Y_L^b \inp{Y_L^a}^\dagger
			\inb{
				  2 Y_L^a \inp{Y_L^b}^\dagger
				+ \Tr\inp{
					3 \inp{Y_Q^b}^\dagger
					Y_Q^a
				+ \inp{Y_L^b}^\dagger
					Y_L^a
			}
		+ 4 \inp{
			\lambdaLight^{cb}}^*
			\lambdaLight^{ca}
			}
\notag	\\
%|%%%%%%%%%%%%%%%%%%%%%%%%%%%%%%%DO NOT EXCEED%%%%%%%%%%%%%%%%%%%%%%%%%%%%%%%|%
	& \quad	\left. {}
		- g_1^2
			\inp{
				  18 f f^\dagger
				+ 3 Y_L^a \inp{Y_L^a}^\dagger
				+ 13 g_1^2
			}
		- 3 g_2^2
			\inp{
				  14 f f^\dagger
				+ Y_L^a \inp{Y_L^a}^\dagger
				+ 3 g_2^2
			}
		+ \text{h.c.}
		\vphantom{\frac{ \abs{F_\phi}^2 }{ \inp{16 \pi^2}^2 }}
		\right]
\label{Eq:Selectron.Mass}
\end{align}
%
%|%%%%%%%%%%%%%%%%%%%%%%%%%%%%%%%DO NOT EXCEED%%%%%%%%%%%%%%%%%%%%%%%%%%%%%%%|%
%
Taking
\begin{equation}
\man = \frac{\Fphi}{16 \pi^2},
\label{Eq:man.def}
\end{equation}
assuming that $f$, $f_c$ are diagonal in flavor space (an assumption required
to satisfy constraints from lepton flavor violating
experiments\cite{Bellgardt:1987du}), and neglecting the first and second
generation yukawa couplings simplifies Eqs.~\eqn{Right.Selectron.Mass} and
\eqn{Selectron.Mass} to
\begin{align}
m_{e^c}^2
	& =	\man^2 \inb{
				  40 f_{c1}^4
				+ 8 f_{c1}^2 \inp{f_{c2}^2 + f_{c3}^2}
				- 48 f_{c1}^2 g_1^2
				- 52 g_1^4
			}
\label{Eq:AMSB.Mass.Selectron.Right}
	\\
m_{e}^2	& =	\man^2 \inb{
				  84 f_{1}^4
				+ 12 f_{1}^2 \inp{f_{2}^2 + f_{3}^2}
				- 6 f_{1}^2 \inp{3 g_1^2 + 7 g_2^2}
				- 13 g_1^4
				- 9 g_2^4
			}
\label{Eq:AMSB.Mass.Selectron.Left}
\end{align}
for the first generation.\footnote{The expressions for the smuon may be gotten 
by taking $f_1 \leftrightarrow f_2$ and $f_{c1} \leftrightarrow f_{c2}$.}
We then only need
\begin{equation}
\label{Eq:f.Limits}
f_1(F_\phi)	\simeq
f_2(F_\phi)	\simeq
f_{c1}(F_\phi)	\simeq
f_{c2}(F_\phi)	\gtrsim 
0.6
\end{equation}
%
%
%|%%%%%%%%%%%%%%%%%%%%%%%%%%%%%%%DO NOT EXCEED%%%%%%%%%%%%%%%%%%%%%%%%%%%%%%%|%
%
to make the sleptons positive (from the detailed analysis of \Sec{Sleptons}).

These couplings and the masses of the doubly-charged field 
and the left-handed triplets
are experimentally constrained from muonium-antimuonium 
oscillations\cite{Willmann:1998gd} 
which demands that
\begin{equation}
\frac{ f_{c1} f_{c2} }{ 4 \sqrt{2} m_{DC}^2 }			\approx
\frac{ f_1 f_2 }{ 4  \sqrt{2} m_{\Delta,\DeltaBar}^2 }		<
3 \E{-3} G_F;
\label{Eq:DC.muonium.constraint}
\end{equation}
%
%|%%%%%%%%%%%%%%%%%%%%%%%%%%%%%%%DO NOT EXCEED%%%%%%%%%%%%%%%%%%%%%%%%%%%%%%%|%
%
The minimum $f$ values that satisfies \eq{f.Limits}
implies a lower bound on the 
masses of the doubly-charged and left-handed triplet Higgs field to be about 
$m_{DC}, m_\Delta \geq 2$ TeV. The lighter end of this range 
is clearly accessible at the \abbr{LHC}.

It is worth noting that even though the $f,f_c$ are diagonal, one may obtain 
large neutrino mixing. As already noted, the neutrino masses arise from the 
type I seesaw\cite{Minkowski:1977sc, Gell-Mann:1980vs, Yanagida:1979as, 
Glashow:1979nm, Mohapatra:1979ia} formula given by:
\begin{align}
\mathcal{M}_{\nu}
	& =	-M^T_D M^{-1}_R M_D
\notag	\\
	& = \frac{v^2_{wk} \sin^2\beta}{v^2_R} y^T_\nu f_c^{-1} y_\nu
\end{align}
Note that the Yukawa coupling matrix $y_\nu$ is arbitrary and can be 
easily arrange to give large mixings even though $f$ is diagonal and we 
can fit the neutrino data by appropriate choice of parameters. 

%|%%%%%%%%%%%%%%%%%%%%%%%%%%%%%%%DO NOT EXCEED%%%%%%%%%%%%%%%%%%%%%%%%%%%%%%%|%

%%%%%%%%%%%%%%%%%%%%%%%%%%%%%%%%%%%%%%%%%%%%%%%%%%%%%%%%%%%%%%%%%%%%%%%%%%%%%%%
%%%%%%%%%%%%%%%%%%%%%%%%%%%%%%%%%%%%%%%%%%%%%%%%%%%%%%%%%%%%%%%%%%%%%%%%%%%%%%%
\section{Between Scales: $v_R$ to $\Fphi$}
\label{Sec:TeV.Model.Spectrum}
%%%%%%%%%%%%%%%%%%%%%%%%%%%%%%%%%%%%%%%%%%%%%%%%%%%%%%%%%%%%%%%%%%%%%%%%%%%%%%%
%%%%%%%%%%%%%%%%%%%%%%%%%%%%%%%%%%%%%%%%%%%%%%%%%%%%%%%%%%%%%%%%%%%%%%%%%%%%%%%

Once $SU(2)_R$ breaks around the seesaw scale of $10^{11}$ GeV, the 
effective theory contains the \abbr{NMSSM}, an extra set
of higgs doublets, a pair of left-handed triplets, and the doubly-charged
fields\footnote{The resulting theory with the additional particle content
might be aptly labeled the \abbr{NMSSM}$++$}.  The non-renormalizable terms of
\eq{SuperW.SUSYLR} also influence the form of the lower scale theory and
produce some important effects that aid in construction of a realistic
low-energy theory.  One significant contribution comes from the higher dimensional 
operators:  
the generation of a \abbr[>-1+ic]{SUSY} mass term for $\Slight$.  Specifically
the terms
\begin{equation}
\frac{ \lambda_{\mathbf{\Slight}}^c }{ \MP \phi }
	  \Tr\inp{\DeltaC \DeltaBarC} \Slight^2
	+ \frac{ \lambda_\Mu }{ \MP \phi } \Sheavy^2 \Slight^2
\label{Eq:SuperW.nr.S.Light.Mass.Terms}
\end{equation}
generate a superpotential term of $\mu_{\Slight} \phi \Slight^2$ when 
$\DeltaC$, $\DeltaBarC$, and $\Sheavy$ get a \abbr{VEV}.  The mass 
$\mu_{\Slight}$ is given by\footnote{We choose to denote the scalar 
component of the superfield $X$ as $\scalar{X}$ to avoid 
confusion between the superfield and its scalar component.  This allows us to
write more meaningful expressions such as $\vev{X}/\vev{\scalar{X}} = \phi$}
\begin{equation}
\mu_{\Slight} 
	\equiv	  \frac{\lambda_{\mathbf{\Slight}}^c }{ \MP } 
			\vev{\scalar{\DeltaC}} \vev{\scalar{\DeltaBarC}}
		+ \frac{ \lambda_\Mu }{ \MP } \vev{\scalar{\Sheavy}}^2
	\simeq \frac{v_R^2}{\MP}
\label{Eq:mu.S.Light}
\end{equation}

%|%%%%%%%%%%%%%%%%%%%%%%%%%%%%%%%DO NOT EXCEED%%%%%%%%%%%%%%%%%%%%%%%%%%%%%%%|%

Because the $v_R$ threshold preserves the \abbr{AMSB} trajectory, this
explicit mass term produces a \abbr{SUSY} breaking bilinear term proportional
to $\Fphi$
\begin{equation}
\intOp[2]{\theta}{ \mu_{\Slight} \phi \Slight^2 }
	\supset \mu_{\Slight} \Fphi \scalar{\Slight}^2
	\equiv b_{\Slight} \scalar{\Slight}^2
\end{equation}
with $b_{\Slight}$ given as
\begin{equation}
b_{\Slight} = \mu_{\Slight} \Fphi \simeq \frac{v_R^2}{\MP} \Fphi.
\label{Eq:bilinearB.S.light}
\end{equation}
In \Sec{Low.Scale.Theory} this term will be shown to play an important role
in \abbr{EWSB}; for now it suffices to note that if $b_{\Slight}$ is to be 
of the
expected order of $\MSUSY^2$, then the right-handed scale must be around 
$v_R \simeq 10^{11}$ GeV.  Constraining $v_R$ automatically
determines the scale of new physics $\MNP$ from \eq{new.physics.expression}:
$\MNP \lesssim 10^{16}$--$10^{18}$ GeV.  The end result 
is that the order of magnitude of
all the scales of the theory are fixed.

%|%%%%%%%%%%%%%%%%%%%%%%%%%%%%%%%DO NOT EXCEED%%%%%%%%%%%%%%%%%%%%%%%%%%%%%%%|%

Furthermore, the non-renormalizable terms can also be used to simplify the
low-energy theory, though this is not necessary.  Consider the terms
\begin{equation}
\frac{\inp{\lambda_\sigma^c}^{ab}}{\MP \phi}
	\Tr\inp{\DeltaC \DeltaBarC}
	\Tr\inp{ \Phi_a \tau_2 \Phi_b^T \tau_2 }
+ \frac{2\lambda_\alpha^c \epsilon^{ab}}{\MP \phi}
	\Tr\inp{\DeltaC \tau_2 \Phi_a^T \tau_2 \Phi_b \DeltaBarC}
\end{equation}
which yield a low energy mass matrix for the $\Phi$'s that is not symmetric
between $\Phi_1$ and $\Phi_2$ (due to the second term).  The asymmetry
 generates an operator of the form:
\begin{equation}
W \supset \I M H_{u2} \tau_2 H_{d1}
\end{equation}
without the corresponding $H_{u1} H_{d2}$ term.  This allows a large mass,
say of order $\Fphi$, for $H_{u2}$ and $H_{d1}$
while leaving $H_{u1}$ and $H_{d2}$ light.  The resulting \abbr{VEV}s for 
$\scalar{H_{u2}}$ and $\scalar{H_{d1}}$ will then be suppressed by $M$ 
and will not play a role in the theory below $\Fphi$.  

%|%%%%%%%%%%%%%%%%%%%%%%%%%%%%%%%DO NOT EXCEED%%%%%%%%%%%%%%%%%%%%%%%%%%%%%%%|%

Finally, as discussed in \Sec{SUSYLR+AMSB.Review}, the non-renormalizable terms
yield masses around $\Fphi$ for the left-handed triplets as well as the
doubly-charged fields.  These fields therefore decouple from the
electroweak scale theory along with the extra bi-doublet due to the doublet-doublet 
splitting mechanism discussed above.  This
leaves the low energy theory as the \abbr{NMSSM} and we use this to explore 
electroweak symmetry breaking as well as the remaining consequences of the low-energy 
theory.

%|%%%%%%%%%%%%%%%%%%%%%%%%%%%%%%%DO NOT EXCEED%%%%%%%%%%%%%%%%%%%%%%%%%%%%%%%|%

\noscalarunderline

%%%%%%%%%%%%%%%%%%%%%%%%%%%%%%%%%%%%%%%%%%%%%%%%%%%%%%%%%%%%%%%%%%%%%%%%%%%%%%%
%%%%%%%%%%%%%%%%%%%%%%%%%%%%%%%%%%%%%%%%%%%%%%%%%%%%%%%%%%%%%%%%%%%%%%%%%%%%%%%
\section{EWSB}
\label{Sec:Low.Scale.Theory}
%%%%%%%%%%%%%%%%%%%%%%%%%%%%%%%%%%%%%%%%%%%%%%%%%%%%%%%%%%%%%%%%%%%%%%%%%%%%%%%
%%%%%%%%%%%%%%%%%%%%%%%%%%%%%%%%%%%%%%%%%%%%%%%%%%%%%%%%%%%%%%%%%%%%%%%%%%%%%%%

Naively it would be expected that the resulting low-energy theory is merely
the \abbr{NMSSM} (since the remaining particle content is precisely that
theory), but if this were the case, the model would not be able to achieve 
a realistic mass spectrum---the singlet $\Slight$ would get a very small
\abbr{VEV}, and the Higgsino would be lighter than allowed by 
experiment\cite{Kitano:2004zd}.  The origin of this problem is best 
illustrated with a toy model:

%|%%%%%%%%%%%%%%%%%%%%%%%%%%%%%%%DO NOT EXCEED%%%%%%%%%%%%%%%%%%%%%%%%%%%%%%%|%

%%%%%%%%%%%%%%%%%%%%%%%%%%%%%%%%%%%%%%%%%%%%%%%%%%%%%%%%%%%%%%%%%%%%%%%%%%%%%%%
% % % % % % % % % % % % % % % % % % % % % % % % % % % % % % % % % % % % % % % %
\subsection{Toy Exposition}
\label{Sec:Low.Scale.Theory.Toy}
% % % % % % % % % % % % % % % % % % % % % % % % % % % % % % % % % % % % % % % %
%%%%%%%%%%%%%%%%%%%%%%%%%%%%%%%%%%%%%%%%%%%%%%%%%%%%%%%%%%%%%%%%%%%%%%%%%%%%%%%

%|%%%%%%%%%%%%%%%%%%%%%%%%%%%%%%%DO NOT EXCEED%%%%%%%%%%%%%%%%%%%%%%%%%%%%%%%|%

Consider a superpotential given by
\begin{equation}
W_{\text{toy}} = \frac{1}{3} \kappa \Slight^3
\label{Eq:AMSB.NMSSM.Toy.SuperW}
\end{equation}
where $\Slight$ is a singlet field with no gauge symmetries.  The resulting
scalar potential, including \abbr{SUSY} breaking, is
\begin{equation}
\label{Eq:AMSB.NMSSM.Toy}
V_{\text{toy}}
	=	  \kappa^2 \abs{\scalar{\Slight}}^4
		+ \frac{1}{3} \inp{
			  a_{\kappa} \scalar{\Slight}^3
			+ a_{\kappa}^* \scalar{\Slight}^{* 3}
			}
		+ m_{\Slight}^2 \abs{\scalar{\Slight}}^2.
\end{equation}

%|%%%%%%%%%%%%%%%%%%%%%%%%%%%%%%%DO NOT EXCEED%%%%%%%%%%%%%%%%%%%%%%%%%%%%%%%|%

Assuming the parameters $\kappa$, $a_\kappa$, and $\vev{\Slight}$ are real, 
the minimization condition for \eq{AMSB.NMSSM.Toy} is
\begin{equation}
2 \kappa^2 \vev{\Slight}^2 + a_\kappa \vev{\Slight} + m_{\Slight}^2 = 0
\label{Eq:AMSB.NMSSM.Toy.min.condition}
\end{equation}
and the solution is given as
\begin{equation}
\vev{\Slight}
	= \frac	{-a_\kappa \pm \sqrt{a_\kappa^2 - 8 \kappa^2 m_{\Slight}^2}}
		{2 \kappa^2}
\label{Eq:AMSB.NMSSM.Toy.S.light.vev}
\end{equation}

%|%%%%%%%%%%%%%%%%%%%%%%%%%%%%%%%DO NOT EXCEED%%%%%%%%%%%%%%%%%%%%%%%%%%%%%%%|%

The soft couplings $a_\kappa$ and $m_{\Slight}$ are determined by \abbr{AMSB}
via Eqs.~\eqn{AMSB.scalar.mass} and \eqn{AMSB.trilinear.A}:
\begin{equation}
\label{Eq:AMSB.NMSSM.Toy.Terms}
\begin{aligned}
a_{\kappa}
	& = \frac{F_{\phi}}{16 \pi^2} 6 \kappa^3
	\\
m^2_{\Slight}
	& = \frac{\abs{F_{\phi}}^2}{\inp{16 \pi^2}^2} 12 \kappa^4
\end{aligned}
\end{equation}

%|%%%%%%%%%%%%%%%%%%%%%%%%%%%%%%%DO NOT EXCEED%%%%%%%%%%%%%%%%%%%%%%%%%%%%%%%|%
%
Substituting these into \eq{AMSB.NMSSM.Toy.S.light.vev} yields
\begin{equation}
\label{Eq:AMSB.NMSSM.Toy.Solution}
\vev{\scalar{\Slight}}
	= \frac{F_{\phi}}{16 \pi^2} \frac{\kappa}{4} \inp{-6 \pm \sqrt{-60}}
\end{equation}
and the large negative under the radical demonstrates the inability to
achieve a real, non-zero \abbr{VEV} in this model.  

%|%%%%%%%%%%%%%%%%%%%%%%%%%%%%%%%DO NOT EXCEED%%%%%%%%%%%%%%%%%%%%%%%%%%%%%%%|%

The source of the problem can be identified by examining the potential of 
$\Slight$.  To expose the difficulty, it is helpful to define
\begin{equation}
x \equiv \frac{\kappa \vev{\Slight}}{\man}
\end{equation} 
and re-write \eq{AMSB.NMSSM.Toy} as
\begin{equation}
\frac{\vev{V_{\text{toy}}}}{4 \man^4}
	= \frac{1}{4 \kappa^2} x^4 + x^3 + 3 \kappa^2 x^2
\end{equation}
where the \abbr{AMSB} expressions of \eq{AMSB.NMSSM.Toy.Terms} have been
substituted.  For the potential to have a non-trivial minimum, it is
necessary that the cubic term dominate for some value of $x$ (since this term
is the only one that provides a negative contribution to the potential);
however, for
large $\kappa$, the $x^2$ term will always be larger than the cubic term.
Meanwhile, for small $\kappa$ the quartic term will dominate the expression.
Therefore, if there is any chance for the $x^3$ term to create a minimum
other than zero, it must be that $\kappa \simeq 1$.  This leaves the potential
as
\begin{equation}
\frac{\vev{V_{\text{toy}}}}{4 \man^4}
	= \frac{1}{4} x^4 + x^3 + 3 x^2
\end{equation}
where it now becomes clear that neither large $x$, $x \sim 1$, nor small $x$ 
will have the cubic term dominate the expression---leaving the only minimum
as the trivial one.  Thus, the heart of the problem is that \abbr{AMSB} 
predicts the cubic term's coefficient such that it will always be weaker 
than either the quartic or quadratic regardless of the parameter regime.

The same problem carries over to the full \abbr{NMSSM}, as pointed out 
in \cite{Kitano:2004zd}.  In this model, the additional coupling of 
$\Slight$ to $H_u$ and $H_d$ does not alter the relative strengths
of $\Slight$'s quartic, cubic, or quadratic terms, but it does add a linear 
term to the potential, $a_\lambda v_u v_d \Slight$.  The induced linear term
shifts the trivial minimum away from zero, but keeps it small.  The 
minimization
condition for $\Slight$ can then be approximated as
\begin{equation}
  \widetilde{\mu}_{\Slight}^2 \vev{\Slight}
- \frac{1}{2 \sqrt2} a_\lambda v^2 \sin 2\beta
= 0
\end{equation}
with $\widetilde{\mu}_{\Slight}^2 \simeq \man^2$ being essentially the
\abbr{AMSB} predicted soft \abbr{SUSY} breaking mass for $\Slight$.  The
maximum value occurs when $\sin 2\beta = 1$ so we have that
\begin{equation}
\vev{\Slight}
	\lesssim \frac{a_\lambda v^2}{2 \widetilde{\mu}_{\Slight}^2 \sqrt2}
	\simeq	 \frac{1}{2 \sqrt 2} \frac{v^2}{\man}
	\simeq	 22 \text{ Gev}
\end{equation}
The small $\vev{\Slight}$ then results in a chargino mass which falls below 
the LEP II bound of about $94$ GeV.

%|%%%%%%%%%%%%%%%%%%%%%%%%%%%%%%%DO NOT EXCEED%%%%%%%%%%%%%%%%%%%%%%%%%%%%%%%|%

Given this limitation of the \abbr{NMSSM}, it is desirable to explore methods
that either alter the relative strengths of the terms or yield a large
tadpole term for $\Slight$.  The former may be done by adding vector-like
matter (as in \cite{Chacko:1999am}), while the latter was explored in
\cite{Kitano:2004zd} by introducing a linear term for $\Slight$.  We propose
here a different solution that alters the relative strengths and is already 
present in the model.

%|%%%%%%%%%%%%%%%%%%%%%%%%%%%%%%%DO NOT EXCEED%%%%%%%%%%%%%%%%%%%%%%%%%%%%%%%|%

%%%%%%%%%%%%%%%%%%%%%%%%%%%%%%%%%%%%%%%%%%%%%%%%%%%%%%%%%%%%%%%%%%%%%%%%%%%%%%%
% % % % % % % % % % % % % % % % % % % % % % % % % % % % % % % % % % % % % % % %
\subsection{Low Energy Theory}
\label{Sec:Low.Scale.Theory.Model}
% % % % % % % % % % % % % % % % % % % % % % % % % % % % % % % % % % % % % % % %
%%%%%%%%%%%%%%%%%%%%%%%%%%%%%%%%%%%%%%%%%%%%%%%%%%%%%%%%%%%%%%%%%%%%%%%%%%%%%%%

%|%%%%%%%%%%%%%%%%%%%%%%%%%%%%%%%DO NOT EXCEED%%%%%%%%%%%%%%%%%%%%%%%%%%%%%%%|%

The superpotential of \eq{SuperW.SUSYLR} contains in its non-renormalizable 
terms the key to solving the small $\vev{\Slight}$ problem: as
discussed in \Sec{TeV.Model.Spectrum}, the terms of \eq{SuperW.SUSYLR.nr.gsv}
generate a mass term for $\Slight$ given by \eq{mu.S.Light}.  This mass term then yields 
a \abbr{SUSY} 
breaking bilinear term given by \eq{bilinearB.S.light}.  The size of 
$b_{\Slight}$ is quite conveniently around the \abbr{SUSY} breaking scale 
and also provides a means of turning the net mass-square of $\Slight$ 
negative.  To establish this property we now turn to the effective 
$\MSUSY$-scale theory.

%|%%%%%%%%%%%%%%%%%%%%%%%%%%%%%%%DO NOT EXCEED%%%%%%%%%%%%%%%%%%%%%%%%%%%%%%%|%

The effective superpotential responsible for \abbr{EWSB} (valid for $\MSUSY 
< Q \ll \Fphi$) is
\begin{align}
\left. W \vphantom{W^T}\right|_{\MSUSY}
	& =	  \I y_{u} Q^T \tau_2 H_u u^c
		+ \I y_{d} Q^T \tau_2 H_d d^c
		+ \I y_{e} L^T \tau_2 H_d e^c
\notag	\\
	& \quad {}
		+ \I \lambda \Slight H_u^T \tau_2 H_d
		+ \half \mu_{\Slight} \Slight^2
		+ \third \kappa \Slight^3
\label{Eq:MSUSY.Super.Potential}
\end{align}
and the \abbr{SUSY} breaking potential is
\begin{align}
\left. V_{\text{SB}} \vphantom{W^T}\right|_{\MSUSY}
	& =	  m_Q^2 \scalar{Q}^\dagger \scalar{Q}
		+ m_{u^c}^2 \scalar{u}^{c \dagger} \scalar{u}^c
		+ m_{d^c}^2 \scalar{d}^{c \dagger} \scalar{d}^c
		+ m_L^2 \scalar{L}^\dagger \scalar{L}
		+ m_{e^c}^2 \scalar{e}^{c \dagger} \scalar{e}^c
%|%%%%%%%%%%%%%%%%%%%%%%%%%%%%%%%DO NOT EXCEED%%%%%%%%%%%%%%%%%%%%%%%%%%%%%%%|%
\notag	\\
	& \quad {}
		+ m_{H_u}^2 \scalar{H_u}^\dagger \scalar{H_u}
		+ m_{H_d}^2 \scalar{H_d}^\dagger \scalar{H_d}
		+ m_{\Slight}^2 \scalar{\Slight}^* \scalar{\Slight}
\notag	\\
%|%%%%%%%%%%%%%%%%%%%%%%%%%%%%%%%DO NOT EXCEED%%%%%%%%%%%%%%%%%%%%%%%%%%%%%%%|%
	& \quad {}
		+ \inb{
			  \I a_{u} \scalar{Q}^T 
			  	\tau_2 \scalar{H_u} \scalar{u^c}
			+ \I a_{d} \scalar{Q}^T 
				\tau_2 \scalar{H_d} \scalar{d^c}
			+ \I a_{e} \scalar{L}^T 
				\tau_2 \scalar{H_d} \scalar{e^c}
			+ \text{h.c.}
		}
%|%%%%%%%%%%%%%%%%%%%%%%%%%%%%%%%DO NOT EXCEED%%%%%%%%%%%%%%%%%%%%%%%%%%%%%%%|%
\notag	\\
	& \quad {}
		+ \inb{
			  \I a_\lambda \scalar{\Slight} 
			  	\scalar{H_u}^T \tau_2 \scalar{H_d}
			- \half b_{\Slight} \scalar{\Slight}^2
			+ \third a_\kappa \scalar{\Slight}^3
			+ \text{h.c.}
		}
%|%%%%%%%%%%%%%%%%%%%%%%%%%%%%%%%DO NOT EXCEED%%%%%%%%%%%%%%%%%%%%%%%%%%%%%%%|%
\notag	\\
	& \quad {}
		- \half \inp{
			  M_3 \lambda_3 \lambda_3
			+ M_2 \lambda_2 \lambda_2
			+ M_1 \lambda_1 \lambda_1
			+ \text{h.c.}
			}
\label{Eq:MSUSY.SUSY.Breaking.Potential}
\end{align}
The resulting Higgs sector potential is
\begin{equation}
V = V_F + V_D + V_{\text{SB}}
\label{Eq:MSUSY.Higgs.Potential}
\end{equation}
with $V_F$ and $V_D$ the typical \abbr{SUSY} contribution:
\begin{align}
V_F	& =	  \abs{\lambda}^2 \abs{\scalar{\Slight}}^2
			\inp{ \abs{\scalar{H_u}}^2 + \abs{\scalar{H_d}}^2 }
		+ \abs{
			  \I \lambda \scalar{H_u}^T \tau_2 \scalar{H_d}
			+ \mu_{\Slight} \scalar{\Slight}
			+ \kappa \scalar{\Slight}^2
			}^2
\label{Eq.MSUSY.Higgs.F.Term.Potential}
	\\
%|%%%%%%%%%%%%%%%%%%%%%%%%%%%%%%%DO NOT EXCEED%%%%%%%%%%%%%%%%%%%%%%%%%%%%%%%|%
V_D	& =	  \eighth \inp{g_1^2 + g_2^2}
			\inp{ \abs{\scalar{H_u}}^2 - \abs{\scalar{H_d}}^2 }^2
		+ \half g_2^2 \abs{\scalar{H_u}^\dagger \scalar{H_d}}^2
\label{Eq.MSUSY.Higgs.D.Term.Potential}
\end{align}

The potential of \eq{MSUSY.Higgs.Potential} can be made to spontaneously 
break electroweak symmetry giving
\begin{align}
\vev{\scalar{H_u}}
	& = \frac{1}{\sqrt2} \begin{pmatrix} 0 \\ v_u \end{pmatrix}	&
\vev{\scalar{H_d}}
	& = \frac{1}{\sqrt2} \begin{pmatrix} v_d \\ 0 \end{pmatrix}	&
\vev{\scalar{\Slight}}
	& = \frac{\Slightvev}{\sqrt2}
\label{Eq:MSUSY.Higgs.VEV.Definitions}
\end{align}
and we take the usual definitions: $v_u = v \sin{\beta}$ and 
$v_d = v \cos{\beta}$.  
The minimization conditions are
\begin{align}
  m_{H_u}^2
- \eighth \inp{g_2^2 + g_1^2} v^2 \cos 2\beta
+ \half \lambda^2 \inp{ \Slightvev^2 + v^2 \cos^2 \beta }
- \frac{\Slightvev}{\sqrt2} \inp{
				  \tilde{a}_\lambda
				+ \frac{\lambda \kappa \Slightvev}{\sqrt2}
				} \cot \beta
	& = 0
\label{Eq:MSUSY.Hu.min.condition}
	\\
%|%%%%%%%%%%%%%%%%%%%%%%%%%%%%%%%DO NOT EXCEED%%%%%%%%%%%%%%%%%%%%%%%%%%%%%%%|%
  m_{H_d}^2
+ \eighth \inp{g_2^2 + g_1^2} v^2 \cos 2\beta
+ \half \lambda^2 \inp{ \Slightvev^2 + v^2 \sin^2 \beta }
- \frac{\Slightvev}{\sqrt2} \inp{
				  \tilde{a}_\lambda
				+ \frac{\lambda \kappa \Slightvev}{\sqrt2}
				} \tan \beta
	& = 0
\label{Eq:MSUSY.Hd.min.condition}
	\\
%|%%%%%%%%%%%%%%%%%%%%%%%%%%%%%%%DO NOT EXCEED%%%%%%%%%%%%%%%%%%%%%%%%%%%%%%%|%
  \tilde{m}_{\Slight}^2
+ \kappa^2 \Slightvev^2
+ \half \lambda^2 v^2
+ \frac{\Slightvev \tilde{a}_\kappa}{\sqrt2}
- \half v^2
	\inp{
		  \frac{\tilde{a}_\lambda}{\Slightvev \sqrt2}
		+ \lambda \kappa
	} \sin 2 \beta
	& = 0
\label{Eq:MSUSY.S.light.min.condition}
\end{align}
The tilded variables are introduced to display the deviations from the 
usual \abbr{NMSSM} due to the presence of the term $\mu_{\Slight}$ in 
\eq{MSUSY.Super.Potential}.  These constructs are defined as
\begin{align}
\tilde{a}_\lambda
	& \equiv		a_\lambda + \lambda \mu_{\Slight}
\label{Eq:MSUSY.a.lambda.S.light.eff}
	\\
\tilde{a}_\kappa
	& \equiv		a_\kappa + 3 \kappa \mu_{\Slight}
\label{Eq:MSUSY.a.kappa.S.light.eff}
	\\
\tilde{m}_{\Slight}^2
	& \equiv		m_{\Slight}^2 + \mu_{\Slight}^2 - b_{\Slight}
\label{Eq:MSUSY.mSqr.S.light.eff}
\end{align}

%|%%%%%%%%%%%%%%%%%%%%%%%%%%%%%%%DO NOT EXCEED%%%%%%%%%%%%%%%%%%%%%%%%%%%%%%%|%

Of particular interest is \eq{MSUSY.mSqr.S.light.eff}, which may be recast
using \eq{bilinearB.S.light} of \Sec{TeV.Model.Spectrum}:
\begin{align}
\tilde{m}_{\Slight}^2
	& =	m_{\Slight}^2 + \mu_{\Slight}^2 - \mu_{\Slight} \Fphi
\notag	\\
	& \approx
		m_{\Slight}^2 - \mu_{\Slight} \Fphi
\notag	\\
	& \simeq
		\inp{
			  \frac{\lambda^4}{\inp{16\pi^2}^2} \Fphi
			- \mu_{\Slight}
		} \Fphi
\notag
\end{align}

%|%%%%%%%%%%%%%%%%%%%%%%%%%%%%%%%DO NOT EXCEED%%%%%%%%%%%%%%%%%%%%%%%%%%%%%%%|%

The second line follows from the fact that
$
\mu_\Slight 
	\sim \mathcal{O}\inp{ \frac{\MSUSY^2}{\Fphi} } 
	\sim \mathcal{O}\inp{ \frac{\Fphi}{\inp{16 \pi^2}^2}  }
$ and therefore
the $\mu_{\Slight}^2$ term is negligible compared to the the other terms. The
last line uses the \abbr{AMSB} expression for the scalar mass-squared, 
assuming
it is dominated by the $\lambda$ contribution.  As can be
seen, due to the $\lambda^4$ suppression, it is relatively easy to adjust
$\mu_{\Slight}$ to the appropriate value to make $\tilde{m}_{\Slight}^2$
negative and therefore induce a singlet \abbr{VEV} of the correct size.  
Given that 
$\lambda(\MSUSY) \lesssim 0.5$ (from constraints of perturbativity to the 
right-handed scale) and that $\mu = \frac{\lambda \Slightvev}{\sqrt{2}}$, 
it is only necessary for $\Slightvev \gtrsim 300$ 
GeV to achieve chargino masses above the LEP II bound.
\begin{figure}[ht]
	\begin{center}
		\begin{picture}(288,180)
			\put(0,0){\includegraphics[scale=1]{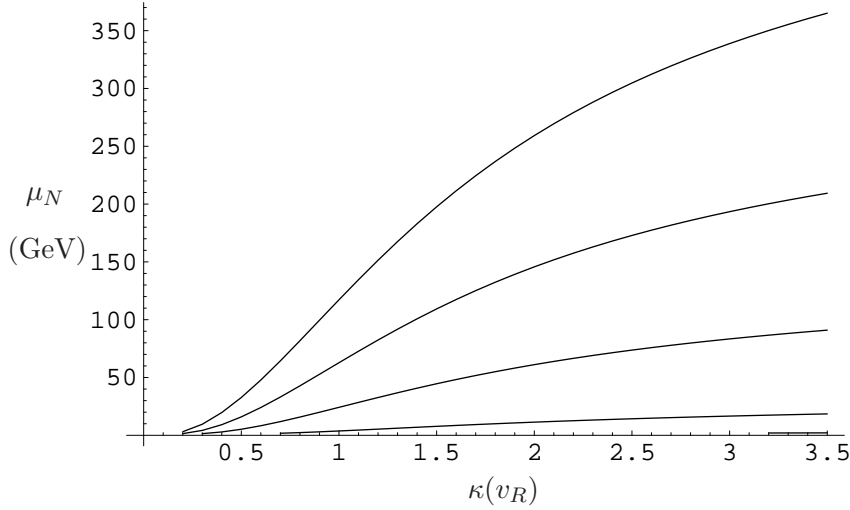}}
			\put(144, -10){$\kappa(v_R)$}
			\put(-30, 90){\parbox{1cm}{$\mu_N$ (GeV)}}
		\end{picture}

	\end{center}
	\caption
	{
		Constant $\Slightvev$ contours in the $\mu_\Slight$-$\kappa(v_R)$ plane where the curves, 
		from top to bottom, correspond to $n = -10000, -7500, -5000, -2500$ and 
		$-1000$ GeV.  A constant value of $\tan{\beta} = 3.25$ has been assumed 
		with $F_\phi = 33$ TeV and $\lambda(v_R) = 0.5$.
	}
	\label{Fig:S.Light.Values}
\end{figure}
\fig{S.Light.Values} shows that such values are easily attainable in this 
situation.  In the figure, constant $\Slightvev$ contours 
are plotted in the $\mu_{\Slight}$--$\kappa(v_R)$ plane treating
the \abbr{VEV}s of the Higgs 
doublets as constant background values with $\tan{\beta} = 3.25$, 
$F_\phi = 33$ TeV, and $\lambda(v_R) = 0.5$.  The ample parameter space
therefore demonstrates that this inherent property of our model easily
provides a means to resolve the conflict between \abbr{AMSB} and the 
\abbr{NMSSM}.

%|%%%%%%%%%%%%%%%%%%%%%%%%%%%%%%%DO NOT EXCEED%%%%%%%%%%%%%%%%%%%%%%%%%%%%%%%|%

The resulting mass spectrum for this \~{N}MSSM is quite similar to the 
\abbr{NMSSM} (see \cite{Ellis:1988er})---particularly for the scalar and 
charged Higgses\footnote{Simply substitute the appropriate variables with 
their tilded form: 
$(a_\kappa, a_\lambda, m_\Slight^2) \rightarrow (\tilde a_\kappa, \tilde a_\lambda, 
\tilde m_\Slight^2)$ Typically, however, $\mu_\Slight$ is rather small and 
so the untilded variables make a good approximation to the tilded ones.}  
whose mass matrices are given by  
\begin{multline}
M_{S}^2	=	
		\\	
		\begin{pmatrix}
			\frac{v_u^2}{4} \inp{g_1^2 + g_2^2} + \frac{\Slightvev v_d}{\sqrt{2} v_u} \tilde A_\Lambda
			&
			\frac{v_d v_u}{4}\inp{4 \lambda^2 + g_1^2 + g_2^2} - \frac{\Slightvev}{\sqrt{2}}\tilde A_\Lambda
			&
			\lambda^2 \Slightvev v_u - \frac{v_d}{\sqrt{2}} \tilde a_\lambda - \lambda \kappa v_d \Slightvev
			\\
			\frac{v_d v_u}{4}\inp{4 \lambda^2 + g_1^2 + g_2^2} - \frac{\Slightvev}{\sqrt{2}}\tilde A_\Lambda
			&
			\frac{v_d^2}{4} \inp{g_1^2 + g_2^2} + \frac{\Slightvev v_u}{\sqrt{2} v_d} \tilde A_\Lambda
			&
			\lambda^2 \Slightvev v_d - \frac{v_u}{\sqrt{2}} \tilde a_\lambda - \lambda \kappa v_u \Slightvev
			\\
			\lambda^2 \Slightvev v_u - \frac{v_d}{\sqrt{2}} \tilde a_\lambda - \lambda \kappa v_d \Slightvev
			&
			\lambda^2 \Slightvev v_d - \frac{v_u}{\sqrt{2}} \tilde a_\lambda - \lambda \kappa v_u \Slightvev
			&
			2 \Slightvev^2 \kappa^2 + \frac{\Slightvev}{\sqrt{2}} \tilde a_\kappa + \frac{v_u v_d}{\sqrt{2} \Slightvev} \tilde a_\lambda
		\end{pmatrix}
\end{multline}
and
\begin{align}
M_C^2 = \frac{v^2}{2 v_d v_u} \inb{\sqrt{2} \Slightvev \tilde A_\Lambda + v_d v_u \inp{\half g_2^2 - \lambda^2}};
\end{align}
defining $\tilde A_\Lambda \equiv \tilde a_\lambda + \frac{\lambda \kappa \Slightvev}{\sqrt{2}}$.

On the other hand, the pseudoscalar mass matrix gets a contribution from 
the $b_\Slight$ term which is rather large and typically guarantees that the heavier 
pseudoscalar is mostly singlet.  Its mass matrix is given by:
%
%|%%%%%%%%%%%%%%%%%%%%%%%%%%%%%%%DO NOT EXCEED%%%%%%%%%%%%%%%%%%%%%%%%%%%%%%%|%
%
% <change date="2008-01-29" time="11:45" by="nsetzer>
\begin{multline}
M_{P}^2
	= \\	\begin{pmatrix}
		\frac{1}{\sqrt{2}} \tilde{A}_\Lambda \frac{v^2 \Slightvev}{v_u v_d}
			&
		\frac{v}{\sqrt{2}}\inp{a_\lambda 
			- \lambda \mu_{\Slight}
			- \sqrt{2} \lambda \kappa \Slightvev
			}
			\\
		\frac{v}{\sqrt{2}}\inp{a_\lambda 
			- \lambda \mu_{\Slight}
			- \sqrt{2} \lambda \kappa \Slightvev
			}
			&
		\frac{v_u v_d}{\Slightvev \sqrt2}
			\inp{	  \widetilde{a}_\lambda
				+ 2 \lambda \kappa \Slightvev \sqrt{2}
				}
		- 3 \widetilde{a}_\kappa \frac{\Slightvev}{\sqrt2}
		+ 2 b_{\Slight}
		+ 8 \kappa \mu_{\Slight} \frac{\Slightvev}{\sqrt2}
		\end{pmatrix}
\end{multline}
% </change>
%

%|%%%%%%%%%%%%%%%%%%%%%%%%%%%%%%%DO NOT EXCEED%%%%%%%%%%%%%%%%%%%%%%%%%%%%%%%|%

The neutralino and chargino mass matrices remain similar to the \abbr{NMSSM}, 
and in the bases
$\inp{\tilde B, \widetilde{W}, \tilde H_u, \tilde H_d, \tilde \Slight}$,
$\inp{\widetilde{W}^+, \tilde H_u^+, \widetilde{W}^-, \tilde H_d^-}$ they
are:
\begin{multline}
M_{\chi^0}	=	
		\\	
		\begin{pmatrix}
			M_1
			&
			0
			&
			\MZ \sin{\beta} \sin{\theta_W}
			&
			-\MZ \cos{\beta} \sin{\theta_W}
			&
			0
			\\
			0
			&
			M_2
			&
			- \MZ \sin{\beta} \cos{\theta_W}
			&
			\MZ \cos{\beta} \cos{\theta_W}
			&
			0
			\\
			\MZ \sin{\beta} \sin{\theta_W}
			&
			- \MZ \sin{\beta} \cos{\theta_W}
			&
			0
			&
			- \frac{\lambda}{\sqrt{2}} \Slightvev
			&
			- \frac{\lambda}{\sqrt{2}} v_d
			\\
			- \MZ \cos{\beta} \sin{\theta_W}
			&
			\MZ \cos{\beta} \cos{\theta_W}
			&
			-\frac{\lambda}{\sqrt{2}} \Slightvev
			&
			0
			&
			-\frac{\lambda}{\sqrt{2}} v_u
			\\
			0
			&
			0
			&
			-\frac{\lambda}{\sqrt{2}} v_d
			&
			-\frac{\lambda}{\sqrt{2}} v_u
			&
			\sqrt{2} \kappa \Slightvev + \mu_{\Slight}
		\end{pmatrix}
\end{multline}

%|%%%%%%%%%%%%%%%%%%%%%%%%%%%%%%%DO NOT EXCEED%%%%%%%%%%%%%%%%%%%%%%%%%%%%%%%|%

\begin{equation}
M_{\chi^{\pm}}	=	
		\begin{pmatrix}
			0
			&
			X^T
			\\
			X
			&
			0
		\end{pmatrix};
		\quad
X	=	
		\begin{pmatrix}
			M_2
			&
			\sqrt{2} \MW \sin{\beta}
			\\
			\sqrt{2} \MW \cos{\beta}
			&
			\mu
		\end{pmatrix}		
\end{equation}
respectively.

%%%%%%%%%%%%%%%%%%%%%%%%%%%%%%%%%%%%%%%%%%%%%%%%%%%%%%%%%%%%%%%%%%%%%%%%%%%%%%%
% % % % % % % % % % % % % % % % % % % % % % % % % % % % % % % % % % % % % % % %
\subsection{A Brief Summary of Scales}
\label{Sec:Brief.Scale.Summary}
% % % % % % % % % % % % % % % % % % % % % % % % % % % % % % % % % % % % % % % %
%%%%%%%%%%%%%%%%%%%%%%%%%%%%%%%%%%%%%%%%%%%%%%%%%%%%%%%%%%%%%%%%%%%%%%%%%%%%%%%

\begin{figure}[ht]
\begin{center}
\includegraphics[scale=1]{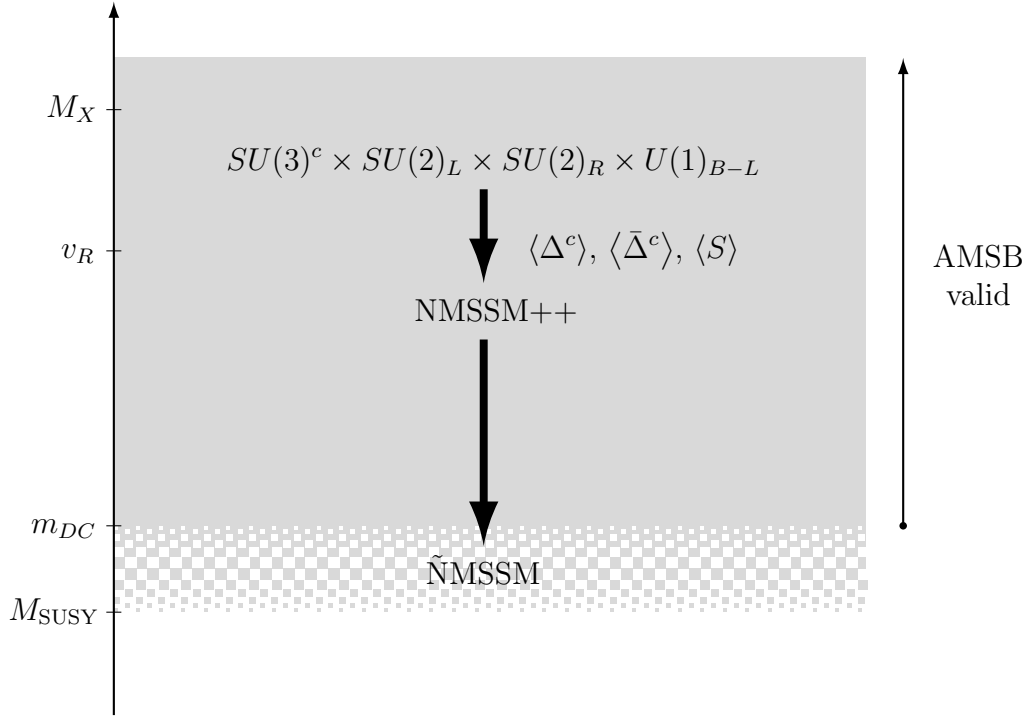}
\end{center}
\caption{A schematic of the SUSYLR$+$AMSB model showing the complete picture
through all the energy scales.}
\label{Fig:SUSYLR+AMSB.schematic}
\end{figure}

With \abbr{EWSB} achieved and the mass spectrum given, we now have a complete 
picture of the physics starting at the high scale $v_R$ and coming down to the
electroweak scale.  The theory starts as a parity-conserving \abbr{SUSYLR}
model with \abbr{AMSB} generating the \abbr{SUSY} breaking, breaks down to
the \abbr{NMSSM}$++$ (but without introducing new \abbr{SUSY} breaking effects),
and finally ends up at $\MSUSY$ as the \~{N}MSSM (as elucidated in
\fig{SUSYLR+AMSB.schematic}).  We may now turn our attention to the rich
phenomenological consequences of this theory.

%
%%%%%%%%%%%%%%%%%%%%%%%%%%%%%%%%%%%%%%%%%%%%%%%%%%%%%%%%%%%%%%%%%%%%%%%%%%%%%%%
%%%%%%%%%%%%%%%%%%%%%%%%%%%%%%%%%%%%%%%%%%%%%%%%%%%%%%%%%%%%%%%%%%%%%%%%%%%%%%%
\section{Phenomenology}
\label{Sec:Sparticle.Masses}
%%%%%%%%%%%%%%%%%%%%%%%%%%%%%%%%%%%%%%%%%%%%%%%%%%%%%%%%%%%%%%%%%%%%%%%%%%%%%%%
%%%%%%%%%%%%%%%%%%%%%%%%%%%%%%%%%%%%%%%%%%%%%%%%%%%%%%%%%%%%%%%%%%%%%%%%%%%%%%%

The following numerical values are based on our parameter running
scheme.  We run
the gauge coupling values from the electroweak
scale to the right-handed scale, $v_R = 2 \times 10^{11}$ 
taking the $F_\phi$ threshold into account by decoupling the triplets 
and doubly charged fields.  Yukawa couplings are then inputs at the 
right-handed scale: the third generation values for the \abbr{SM} couplings 
($y_Q, y_L$) and
all three generations of the seesaw couplings ($f, f_c$).  These
are evolved down to the \abbr{SUSY} scale \cite{Setzer:2005hg, Martin:1993zk}.  
Because of 
parity $f = f_c$ at the right-handed scale and all of the off diagonal
terms are taken to be negligible due to lepton flavor violating constraints.  
We also assume that the first and second generation seesaw
couplings are equal ($f_2 = f_1$) for simplicity.  Soft terms follow their 
\abbr{AMSB} trajectory, given by
Eqs.~\eqn{AMSB.scalar.mass}, \eqn{AMSB.trilinear.A}
and \eqn{AMSB.gaugino.mass} down to the $F_\phi$ scale, below which  the soft 
terms are evolved
to the \abbr{SUSY} scale using the usual \abbr{RGE}s of the 
\abbr{NMSSM} \cite{King:1995vk}.  Note that due to the mass splitting between 
the Higgsinos and Higgses of both the doubly charged and left-handed triplets 
descriped in \Sec{Left-Right}, there will be some corrections to the \abbr{SUSY} 
\abbr{RGE}s.  These corrections will depend on the mass splitting and will 
be fairly small.

%|%%%%%%%%%%%%%%%%%%%%%%%%%%%%%%%DO NOT EXCEED%%%%%%%%%%%%%%%%%%%%%%%%%%%%%%%|%

Numerical results will be compared to popular \abbr{SUSY} 
breaking models: 
\abbr{mSUGRA}, \abbr{mGMSB} and \abbr{mAMSB}---an \abbr{AMSB} in which the 
slepton
mass problem is fixed by adding a universal mass, $m_0$ to all sfermion soft 
masses \cite{Giudice:1998xp}.  Note that slepton phenomenological 
comparisons to \abbr{mAMSB} also apply to  
\cite{Allanach:2000gu} since the 
additional $R$-parity violating lepton sector Yukawa coupling is analogues to 
adding a universal slepton mass.

%%%%%%%%%%%%%%%%%%%%%%%%%%%%%%%%%%%%%%%%%%%%%%%%%%%%%%%%%%%%%%%%%%%%%%%%%%%%%%%
% % % % % % % % % % % % % % % % % % % % % % % % % % % % % % % % % % % % % % % %
\subsection{The Spectrum}
\label{Sec:Spectrum}
% % % % % % % % % % % % % % % % % % % % % % % % % % % % % % % % % % % % % % % %
%%%%%%%%%%%%%%%%%%%%%%%%%%%%%%%%%%%%%%%%%%%%%%%%%%%%%%%%%%%%%%%%%%%%%%%%%%%%%%%

%|%%%%%%%%%%%%%%%%%%%%%%%%%%%%%%%DO NOT EXCEED%%%%%%%%%%%%%%%%%%%%%%%%%%%%%%%|%

\begin{figure}[t]
\begin{center}
\includegraphics[scale=1]{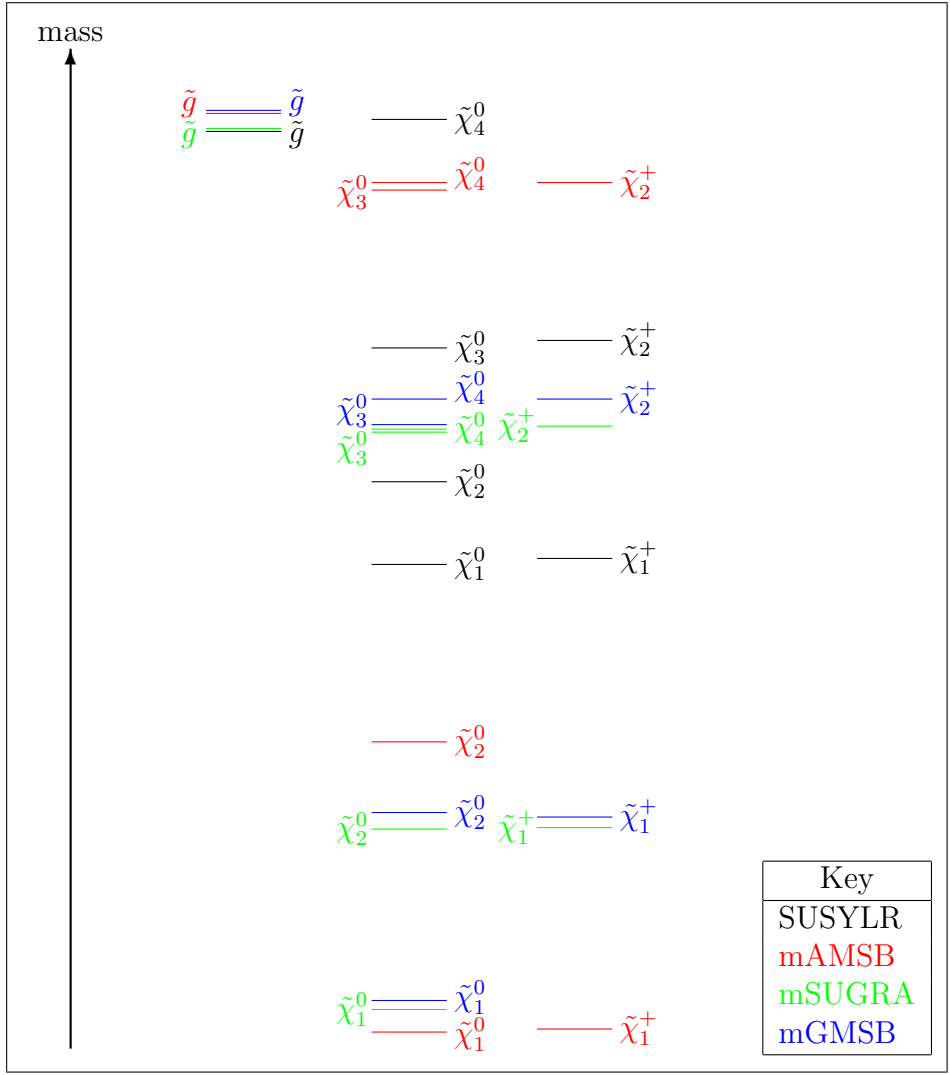}
\end{center}
	\caption
	{
		From left to right, columns correspond to charginos, neutralinos and gluino 
		masses at $\tan{\beta} = 3.25$ and $\sgn{\mu} = +1$.  The parameter points 
		are: $F_\phi = 33$ TeV, $f_1(v_R) = f_3(v_R) = 3.5$ for \thismodel; 
		$m_0 = 209$ GeV, $m_{\frac{1}{2}} = -300$ GeV and 
		$A_0 = 265$ GeV for \abbr{mSUGRA}; $\Lambda = 99$ TeV, $\Mmess = 792$ TeV and 
		$N_5 = 1$ for \abbr{mGMSB}; $F_\phi = 33$ TeV and $m_0 = 645$ GeV for 
		\abbr{mAMSB} (here we also matched to the lightest slepton).
	}
	\label{Fig:particle.spectrum.fermions}
\end{figure}

\begin{figure}[t]
\begin{center}
\includegraphics[scale=1]{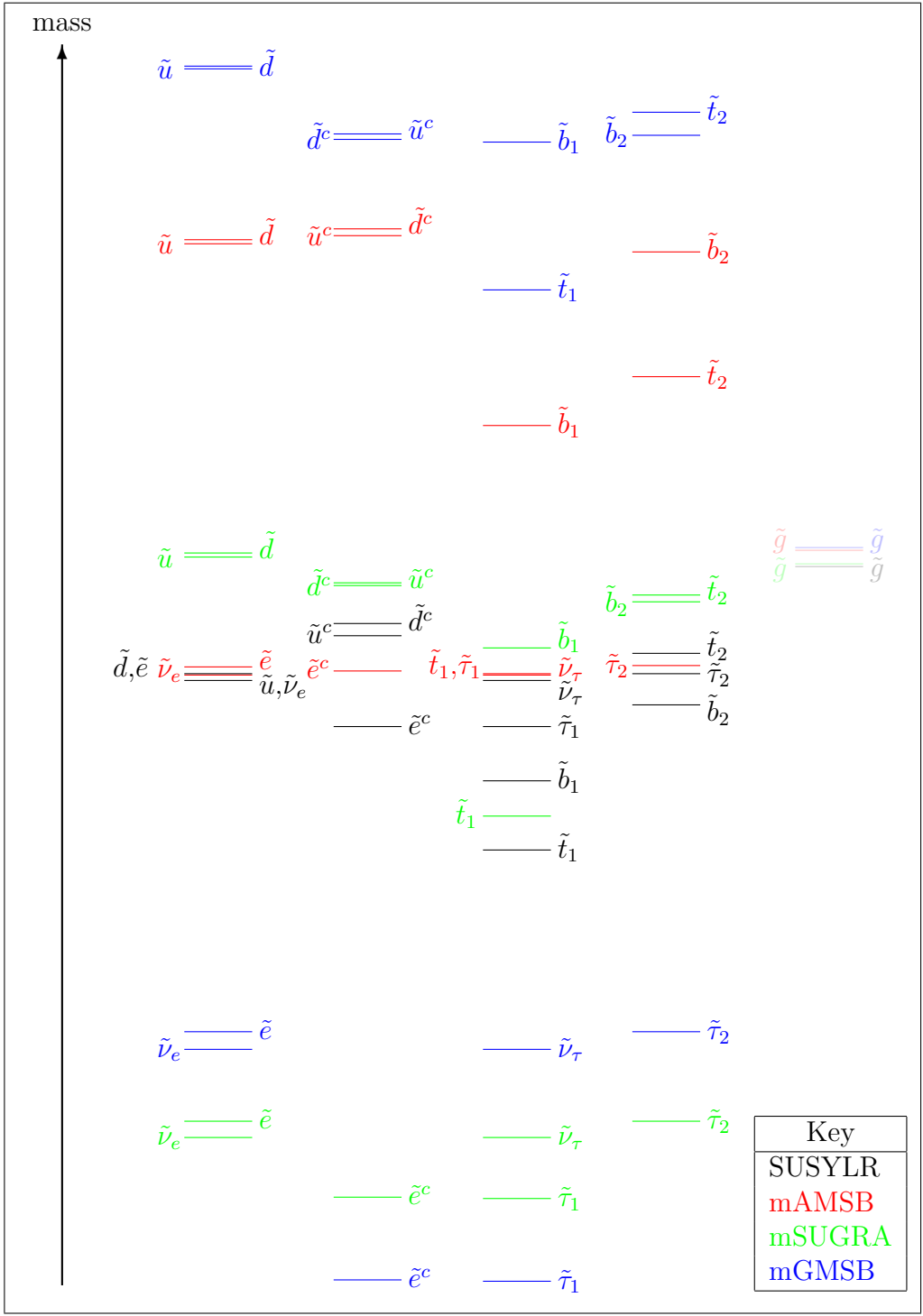}
\end{center}
	\caption
	{
		From left to right, columns correspond to first generation left-handed, 
		first generation right-handed, lightest third generation and heaviest 
		third generation sfermions.  The final column consists of gluino masses 
		for comparison with \fig{particle.spectrum.fermions}.  Input parameters are
		as given in \fig{particle.spectrum.fermions}.
	}
	\label{Fig:particle.spectrum.bosons}
\end{figure}

Before engaging in the full details of the various sectors of the model, it is 
helpful to take a step back and look at the overall spectrum.  
\fig{particle.spectrum.fermions} examines the bosinos and 
\fig{particle.spectrum.bosons} the sfermions 
in this model and compares their masses 
to similar points in parameter space for \abbr{mSUGRA}, \abbr{mGMSB} and 
\abbr{mAMSB} calculated from isajet \cite{Paige:2003mg} 
(matching between the different points were done based on the gluino mass).  
The columns of the bosino chart, \fig{particle.spectrum.fermions}, 
from left to right are gluino, neutralino and chargino.  The columns of the 
sfermion chart, \fig{particle.spectrum.bosons} from left to right are: 
left-handed first generation, right-handed first generation, lightest mass 
third generation (and third generation neutrinos), heaviest third generation 
and gluinos---for comparison with the bosino chart. The Higgses and the 
mostly singlino neutralino 
have not been included to keep from clottering the plots, although their masses 
are reported in \tbl{Spectrum}.

%|%%%%%%%%%%%%%%%%%%%%%%%%%%%%%%%DO NOT EXCEED%%%%%%%%%%%%%%%%%%%%%%%%%%%%%%%|%

%|%%%%%%%%%%%%%%%%%%%%%%%%%%%%%%%DO NOT EXCEED%%%%%%%%%%%%%%%%%%%%%%%%%%%%%%%|%

%|%%%%%%%%%%%%%%%%%%%%%%%%%%%%%%%DO NOT EXCEED%%%%%%%%%%%%%%%%%%%%%%%%%%%%%%%|%

%|%%%%%%%%%%%%%%%%%%%%%%%%%%%%%%%DO NOT EXCEED%%%%%%%%%%%%%%%%%%%%%%%%%%%%%%%|%

\begin{table}[t]
\begin{center}
\begin{tabular}{c|c|c|c}
	Sfermions	&	Masses (GeV)	&	Bosinos and Higges	&	Masses (GeV)
	\\
\hline
	$\tilde u$	&	$623$				&	$\tilde g$	&	$472$
	\\
	$\tilde d$	&	$627$				&	$\tilde \chi_1^0$	&	$417$
	\\
	$\tilde e$	&	$623/484$				&	$\tilde \chi_2^0$	&	$472$
	\\
	$\tilde \nu_e$	&	$621/479$				&	$\tilde \chi_3^0$	&	$561$
	\\
	$\tilde u^c$	&	$654$				&	$\tilde \chi_4^0$	&	$713$
	\\
	$\tilde d^c$	&	$662$				&	$\tilde \chi_5^0$	&	$1644$
	\\
	$\tilde e^c$	&	$587/438$				&	$\tilde \chi_1^+$	&	$421$
	\\
	$\tilde t_1$	&	$496$				&	$\tilde \chi_2^+$	&	$565$
	\\
	$\tilde b_1$	&	$547$				&	$h^0$	&	$116$
	\\
	$\tilde \tau_1$	&	$587/438$				&	$A^0$	&	$518$
	\\
	$\tilde \nu_\tau$	&	$621/479$				&	$H^0$	&	$523$
	\\
	$\tilde t_2$	&	$641$				&	$H^+$	&	$526$
	\\
	$\tilde b_2$	&	$603$				&	$A_2^0$	&	$2086$
	\\
	$\tilde \tau_2$	&	$625/485$				&	$H_3^0$	&	$1284$
	\\
\end{tabular}
\end{center}
\caption{
	Mass spectrum for the \thismodel{} point given in 
	\fig{particle.spectrum.fermions}.  Slepton masses are reported for 
	$f_1(v_R) = f_3(v_R) = 3.5/1.4$.  Higgs masses are also reported here as well as the 
	mostly singlino neutralino.
}
\label{Table:Spectrum}
\end{table}

%|%%%%%%%%%%%%%%%%%%%%%%%%%%%%%%%DO NOT EXCEED%%%%%%%%%%%%%%%%%%%%%%%%%%%%%%%|%

The most striking general feature of these figures is the degeneracy of the 
spectrum between colored and electroweak particles in \thismodel{}.  While 
this is very dependent on the seesaw couplings 
(\tbl{Spectrum} shows slepton masses that are lighter than the squarks due to 
smaller values of the seesaw couplings), it is a possibility that is difficult 
to achieve in other models.  
\tbl{Spectrum} also shows the Higgs masses.  
Here $H_3$ and $A_2$ are the mostly singlet scalar and pseudoscalar.   Due to 
the large size of the singlet \abbr{VEV}, these fields decouple from the 
spectrum as does the mostly singlino $\tilde \chi_5^0$.  The neutral scalar 
Higgs masses stated include the full radiative corrections due to top and stop 
loops \cite{Ellwanger:1993hn}.  These corrections need $m_{\tilde t} \gtrsim 
600$ GeV which implies $F_\phi \gtrsim 33$ TeV to allow the Higgs to be above 
the LEP II bound.  In the following subsections, we will continue 
to explore this spectrum, focusing on the sleptons, squarks and finally the 
neutralinos and charginos.

%|%%%%%%%%%%%%%%%%%%%%%%%%%%%%%%%DO NOT EXCEED%%%%%%%%%%%%%%%%%%%%%%%%%%%%%%%|%

%%%%%%%%%%%%%%%%%%%%%%%%%%%%%%%%%%%%%%%%%%%%%%%%%%%%%%%%%%%%%%%%%%%%%%%%%%%%%%%
% % % % % % % % % % % % % % % % % % % % % % % % % % % % % % % % % % % % % % % %
\subsection{Sleptons}
\label{Sec:Sleptons}
% % % % % % % % % % % % % % % % % % % % % % % % % % % % % % % % % % % % % % % %
%%%%%%%%%%%%%%%%%%%%%%%%%%%%%%%%%%%%%%%%%%%%%%%%%%%%%%%%%%%%%%%%%%%%%%%%%%%%%%%

\begin{figure}[ht]
\begin{center}
	\subfloat[]{
		\begin{picture}(220,130)
		\put(-10, 80){$f_{c1}$}
		\put(90, 0){$\log_{10}{\frac{Q}{1 GeV}}$}
		\put(5,10){	\includegraphics[scale=0.7]%
				{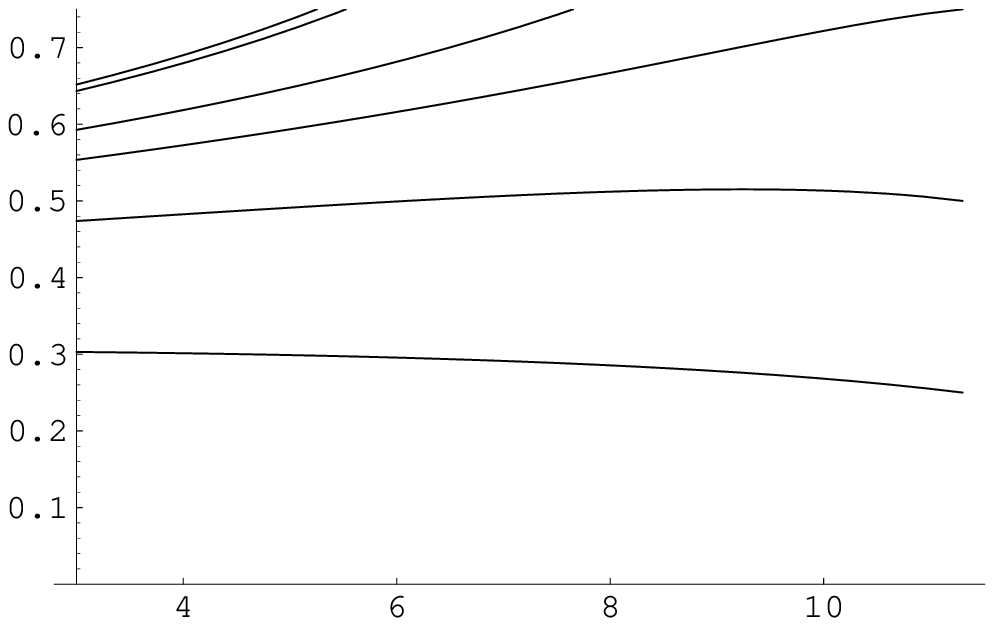}
			}
		\end{picture}
		\label{Fig:f.FIXED.POINT.f3.Zero}
		}
	\subfloat[]{
		\begin{picture}(220,130)
		\put(-10, 80){$f_{c1}$}
		\put(90,0){$\log_{10}{\frac{Q}{1 GeV}}$}
		\put(5,10){	\includegraphics[scale=0.7]%
				{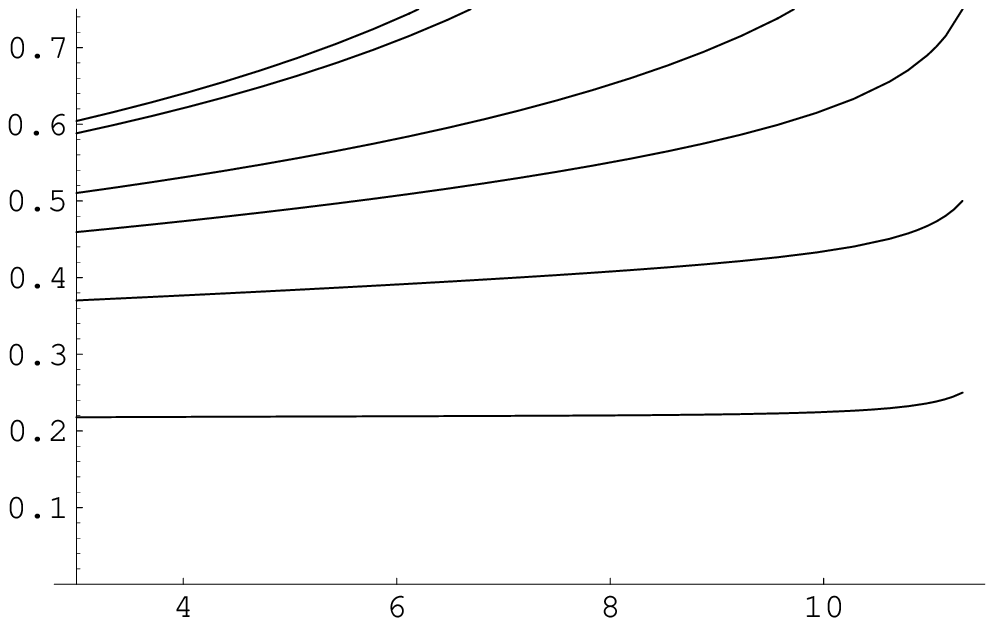}
			}
		\end{picture}
		\label{Fig:f.FIXED.POINT.f3.3.5}
		}
\end{center}
	\caption{
		Plots of $f_{c1}$ verses the log of the energy scale.  The lines
		correspond, in ascending order, to $f_1(v_R)$  values of
		$0.25$, $0.5$, $0.75$, $1$,
		$2.25$ and $3.5$ for 
		\protect\subref{Fig:f.FIXED.POINT.f3.Zero} $f_3(v_R) = 0$ and 
		\protect\subref{Fig:f.FIXED.POINT.f3.3.5} $f_3(v_R) = 3.5$.
		}
\label{Fig:f.FIXED.POINT}
\end{figure}

We start this discussion by analyzing the seeseaw yukawa
couplings $f$ and $f_c$.   In all the work that follows, we take their 
maximum value at $v_R$ to be $\sim \sqrt{4 \pi} \sim 3.5$ based on 
perturbativity arguments.  
Of immediate note is the fixed point-like
behavior of these couplings.  This can be seen in \fig{f.FIXED.POINT}, 
which plots $f_{c1}$ verses the
log of the energy scale for initial values
of (a) $f_3(v_R) = 0$ and (b) $f_3(v_R) = 3.5$; the curves,
in ascending order, correspond to $f_1(v_R) = 0.25$, $0.5$, $0.75$,
$1$, $2.25$, $3.5$.  Increasing the initial value of $f_3$ decreases the value
 of
$f_1$ at the TeV scale as can be seen by comparing \fig{f.FIXED.POINT.f3.Zero} 
and \fig{f.FIXED.POINT.f3.3.5}.

%|%%%%%%%%%%%%%%%%%%%%%%%%%%%%%%%DO NOT EXCEED%%%%%%%%%%%%%%%%%%%%%%%%%%%%%%%|%

%|%%%%%%%%%%%%%%%%%%%%%%%%%%%%%%%DO NOT EXCEED%%%%%%%%%%%%%%%%%%%%%%%%%%%%%%%|%

Similar plots can be drawn for the other couplings: $f_1, f_{c3}$ and
$f_{c1}$,
but their qualitative behavior follows those in \fig{f.FIXED.POINT}.
\tbl{Fixed.Point.Values} illustrates the quantitative differences in the values of
the fixed points.  For initial values of $f_1$ and
$f_3$ greater than $1.5$, these values are correct up to $2\%$.  The higher
values
for the right-handed sector ($f_c$) are due to the slower running caused by 
the broken $SU(2)_R$ symmetry.

%|%%%%%%%%%%%%%%%%%%%%%%%%%%%%%%%DO NOT EXCEED%%%%%%%%%%%%%%%%%%%%%%%%%%%%%%%|%

\begin{table}[ht]
\begin{center}
\begin{tabular}{c|c|c|c|c}
	&	$\quad f_3 \quad$
	&	$\quad f_1 \quad$
	&	$\quad f_{c3} \quad$
	&	$\quad f_{c1} \quad$
	\\
\hline
Fixed Point Value
	&	$0.64$  % f_3
	&	$0.64$  % f_1
	&	$0.67$  % f_{c3}
	&	$0.67$  % f_{c1}
\end{tabular}
\end{center}
\caption{
Fixed point values at the DC scale for the seesaw couplings
assuming initial values are above $1.5$ for the data point used in 
\fig{f.FIXED.POINT}.
}
\label{Table:Fixed.Point.Values}
\end{table}

%|%%%%%%%%%%%%%%%%%%%%%%%%%%%%%%%DO NOT EXCEED%%%%%%%%%%%%%%%%%%%%%%%%%%%%%%%|%

\begin{figure}[ht]
	\begin{center}
		\begin{picture}(288,180)
			\put(0,0){\includegraphics[scale=1]{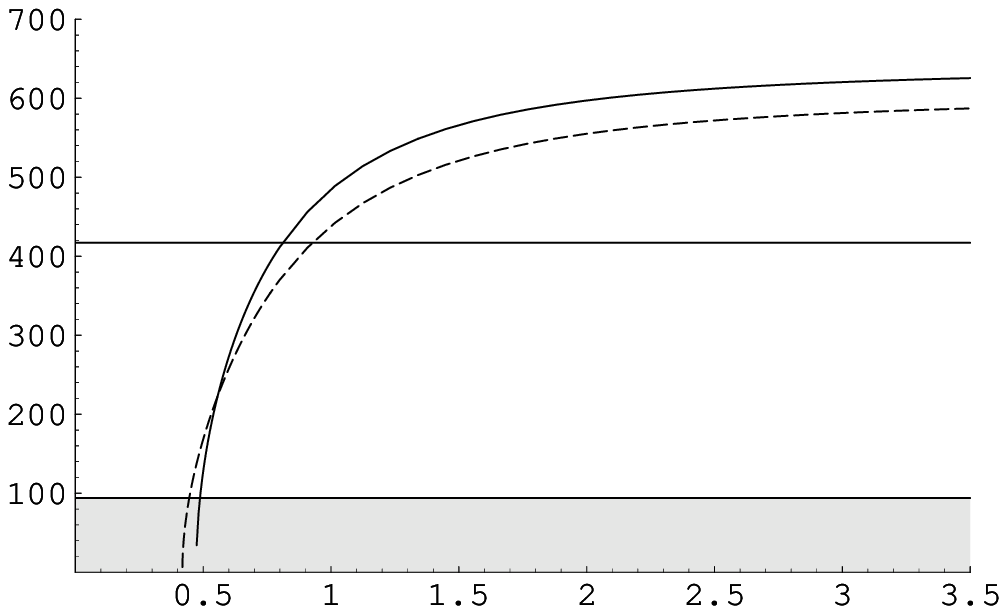}}
			\put(30, 115){$m_{\tilde{\chi}_1}$} %mass of the LSP
			\put(-35, 90){\parbox{1cm}{Mass (GeV)}}
			\put(144, -10){$f(v_R)$}
		\end{picture}

	\end{center}
	\caption{
		Plot of $m_{\tilde{e}^c}$ (dashed) and $m_{\tilde{e}}$ as a
		function of $f_1(v_R) = f_3(v_R)$ for $F_\phi = 33$ TeV.  The
		greyed-out region has been excluded by LEP II and the line at
		around $417$ GeV is the mass of the neutralino, the \abbr{LSP} 
		in this
		case.
		}
	\label{Fig:SLEPTON.MASS.VERSES.f}
\end{figure}

This fixed point like behavior translates into an upper bound for the
slepton masses.  This can be seen in
\fig{SLEPTON.MASS.VERSES.f}, which displays the dependence of the selectron
masses on the
initial value  of the seesaw coupling.  For this
plot $f_1(v_R) = f_3(v_R)$ 
has been assumed for simplicity.  For 
$f \geq 0.5$ the yukawa coupling
contribution is comparable in size to the gauge coupling contribution in the
\abbr{AMSB} mass expression, \textit{e.g.} \eq{Right.Selectron.Mass}.  The 
mass' quartic dependence 
on the seesaw couplings is reflected in its steep rise near $0.5$ 
and its rapid surpassing of the 
LEP II bound.  At a value of $f \sim 1$ this
steep ascent slows down indicating the onset of the fixed point behavior,
beyond which the low energy observable $f(F_\phi)$ values are approximately $0.6$.

%|%%%%%%%%%%%%%%%%%%%%%%%%%%%%%%%DO NOT EXCEED%%%%%%%%%%%%%%%%%%%%%%%%%%%%%%%|%

%|%%%%%%%%%%%%%%%%%%%%%%%%%%%%%%%DO NOT EXCEED%%%%%%%%%%%%%%%%%%%%%%%%%%%%%%%|%

The masses of the other sleptons follow the behavior of
\fig{SLEPTON.MASS.VERSES.f}, a general feature of which is the mild 
degeneracy between the left and right -handed slepton masses.  
This seems a bit contrary to 
Eqs.~\eqn{AMSB.Mass.Selectron.Right} and 
\eqn{AMSB.Mass.Selectron.Left}, which show that the factor for $f_1^4$ 
term for the 
left-handed sleptons is twice as large as that of the right-handed sleptons.  
However, this term is capped by the fixed-point of $f_1$ and the negative 
$SU(2)_L$ contribution happens to be a little less than half of this value 
(an accidental cancelation) yielding the degeneracy.

%|%%%%%%%%%%%%%%%%%%%%%%%%%%%%%%%DO NOT EXCEED%%%%%%%%%%%%%%%%%%%%%%%%%%%%%%%|%

This is an interesting situation phenomologically since it numerically falls 
in between 
\abbr{mSUGRA}/\abbr{mGMSB} and \abbr{mAMSB}.  In \abbr{mSUGRA}, left-handed 
slepton masses get larger positive contributions from $M_2$ as they run 
from the ultraviolet.  In \abbr{mGMSB} 
boundary conditions dictate that the left-handed to right-handed mass ratio 
is about $2:1$.  
Meanwhile, in \abbr{mAMSB}, 
both sectors get the same contribution from $m_0$, the universal masses needed 
to make the sleptons non-tachyonic, which drops out in the mass splitting  
at tree level.  Furthermore, there are accidental cancellations 
in the anomaly induced splittings related to the gauge contributions and in 
the $D$-term contributions \cite{Gherghetta:1999sw, Feng:1999hg}.  
The upshot of this is that the mass splitting is usually dominated by 
loop-level effects and is quite small \cite{Gherghetta:1999sw}.

%|%%%%%%%%%%%%%%%%%%%%%%%%%%%%%%%DO NOT EXCEED%%%%%%%%%%%%%%%%%%%%%%%%%%%%%%%|%

\begin{figure}[ht]
\begin{center}
	\begin{picture}(288,180)
		\put(0,0){\includegraphics[scale=1]{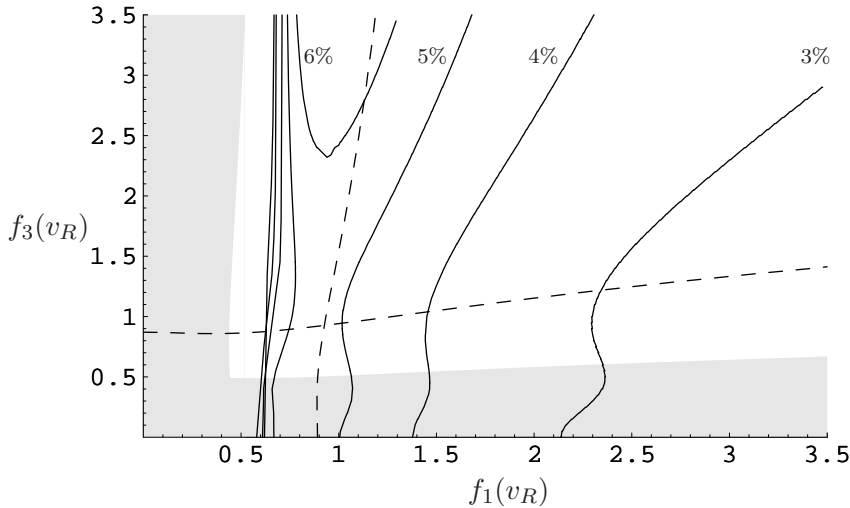}}
		\put(144, -10){$f_1(v_R)$}
		\put(-30, 90){$f_3(v_R)$}
		\put(270, 155){\scriptsize{$3 \%$}}
		\put(167, 155){\scriptsize{$4 \%$}}
		\put(125, 155){\scriptsize{$5 \%$}}
		\put(82, 155){\scriptsize{$6 \%$}}
	\end{picture}
	
\end{center}
\caption
	{
		Constant contours of $\frac{m_{\tilde e} - m_{\tilde e^c}}{m_{\tilde e} +
		 m_{\tilde e^c}} \times 100 \%$ in the $f_3(v_R)-f_1(v_R)$ plane.  The 
		 unlabeled contours on the left side of the plot, from left to right, 
		 correspond to $2 \%, 3 \%, 4 \%$ and $5 \%$.  The dashed vertical 
		 (horizontal) contour corresponds to a $\tilde \tau_1$ ($\tilde e^c$)constant 
		 contour of mass equal to that of the \abbr{LSP} ($417$ GeV).  The shaded region 
		 is excluded by LEP II bounds of $81.9$ GeV ($94$ GeV) on the mass of 
		 $\tilde \tau_1$ ($\tilde e^c$).
	}
\label{Fig:PERCENT.DIFFERENCE.CONTOURS}
\end{figure}

As a concrete example for \abbr{mAMSB}, including the first loop leading log, 
the difference between the masses squared with 
$\tan{\beta} = 3.25$ and $F_\phi = 33$ TeV is given by $ \Delta_e = 
m_{\tilde e_L}^2 - m_{\tilde e_R}^2 \sim 751$ 
$\text{GeV}^2$ \cite{Gherghetta:1999sw, Feng:1999hg}.  The corresponding 
percent difference, defined as 
$\frac{\Delta_e}{(m_{\tilde e_L} + m_{\tilde e_R})^2}$, is then 
highly dependent on the masses of the selectrons.  For selectron massses 
above the mass of the \abbr{LSP} given in \fig{SLEPTON.MASS.VERSES.f}, 
$\sim 450$ GeV, the percent difference is less than $1 \%$.  However, in 
\thismodel{}, the percent difference can rise as high as $5 \%$ as 
demonstrated in \fig{PERCENT.DIFFERENCE.CONTOURS}, which gives contours 
for constant mass percent differences.  Resolution of slepton masses from 
end-point lepton distribtution of the selectron decays at lepton collider is 
roughly $2 \%$ \cite{Danielson:1996tf} making the measurement of such 
mass differences feasible.  Therefore, 
measurements of mild 
mass differences of about $3 - 5\%$ will signle this model out from the large 
mass differences of \abbr{mSUGRA} and \abbr{mGMSB} while potentially 
discriminating it from the small mass differences of \abbr{mAMSB} 
(although this will highly dependent on the values of the seesaw couplings).

%|%%%%%%%%%%%%%%%%%%%%%%%%%%%%%%%DO NOT EXCEED%%%%%%%%%%%%%%%%%%%%%%%%%%%%%%%|%

%|%%%%%%%%%%%%%%%%%%%%%%%%%%%%%%%DO NOT EXCEED%%%%%%%%%%%%%%%%%%%%%%%%%%%%%%%|%

\begin{figure}[ht]
\begin{center}
	\begin{picture}(288,180)
		\put(0,0){\includegraphics[scale=1]{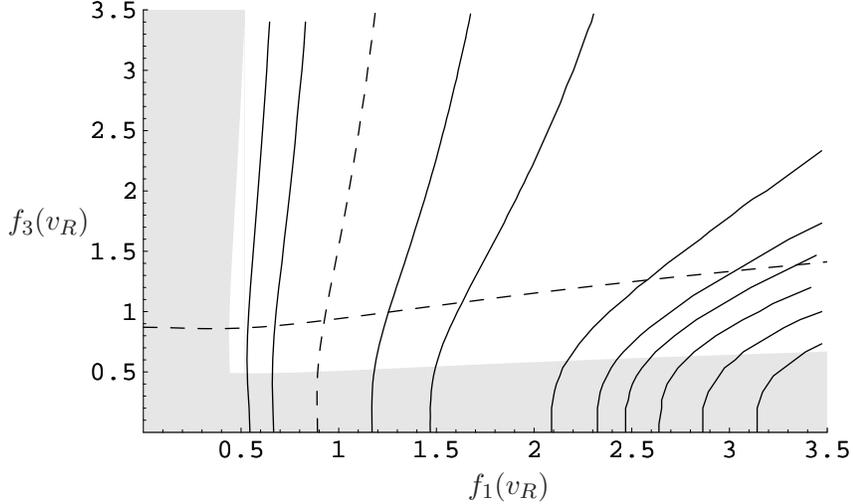}}
		\put(144, -10){$f_1(v_R)$}
		\put(-30, 90){$f_3(v_R)$}
	\end{picture}
\end{center}
\caption{
	Mass contours for the right-handed selectron mass, $m_{\tilde e}$ in 
	the $f_3(v_R)$--$f_1(v_R)$ plane.
	The horizontal and vertical grayed-out contours are ruled out due to 
	LEP II bounds on the lightest stau and selectron masses of $81.9$ GeV 
	and $94$ GeV respectively.  Constant mass contours for the selectron mass 
	$m_{\tilde{e}}= 94$ (the LEP lower bound), $200$, $300$, $417$ 
	(the mass of the lightest neutralino, indicated with a dashed contour) $500$, 
	$550$, $600$, 
	$610$, $615$, $620$, $625$, $630$ GeV, for $F_\phi = 33$ TeV.  The dashed 
	horizontal curve corresponds to a $m_{\tilde \tau_1}$ constant contour equal 
	to the mass of the lightest neutralino.  	The influence of $f_3(v_R)$ is 
	apparent at 	large values of $f_1(v_R)$ and $f_3(v_R)$.  Larger increases in 
	$f_1(v_R)$ are needed for as the mass increases because of the fixed 
	point like behavior of $f_1$.
	}
\label{Fig:SLEPTON.MASS.CONTOUR}
\end{figure}

Constant mass contours for the right-handed selectron are
plotted in \fig{SLEPTON.MASS.CONTOUR} in the $f_3(v_R)$--$f_1(v_R)$ plane.  
This plot allows a study of how the
masses change with respect to both seesaw couplings.  
The horizontal and vertical grayed-out contours 
are ruled out due to LEP II bounds on the lightest stau and selectron masses 
of $81.9$ GeV 
and $94$ GeV respectively.  Mass contours increase from left to right and 
correspond to $m_{\tilde{e}}= 200$, $300$, $417$ (the mass of the lightest 
neutralino, 
indicated with a dashed contour) $500$, $550$, 
$600$, $610$, $615$, $620$, $625$, $630$ GeV.  The horizontal dashed curve 
represents a constant $m_{\tilde \tau_1}$ contour at the mass of the lightest 
neutralino.  Since the selectron is a first generation slepton its mass is 
mainly governed
by $f_1$, \eq{Selectron.Mass}, explaining the small dependence on $f_3$
for smaller values of $f_1$.  Two things are clear from this plot: the fixed
point like behavior---reflected in the fact that for large $f_1$ an equal
change in mass requires a larger change in $f_1$---and the decrease of the
$f_1$ fixed point with the increase of $f_3$.  This latter point is
responsible for the curving to the right of the contours at high $f_1$
values and was mentioned earlier with regards to \fig{f.FIXED.POINT}.

%|%%%%%%%%%%%%%%%%%%%%%%%%%%%%%%%DO NOT EXCEED%%%%%%%%%%%%%%%%%%%%%%%%%%%%%%%|%

%|%%%%%%%%%%%%%%%%%%%%%%%%%%%%%%%DO NOT EXCEED%%%%%%%%%%%%%%%%%%%%%%%%%%%%%%%|%

As a final remark on the sleptons, notice that the contours in 
\fig{SLEPTON.MASS.CONTOUR} correpsonding to the \abbr{LSP} suggest more 
stringent lower bounds on the seesaw couplings than the LEP II bounds.  These 
are necessary so that the lightest neutralino will be the \abbr{LSP} 
and therefore a possible dark matter candidate.  The values indicated in the plot 
correspond to low energy values of the seesaw couplings that are only about 
$10\%$ off from their fixed-point value, $f_{c1}, f_{c3}, f_{1}, f_{3} 
\sim 0.6$.  Therefore, for succesful dark matter, the seesaw couplings can 
be expected to be larger than about $0.5$.  This can be checked by a quick 
calculation, since the lightest neutralino mass is approximately the wino mass 
(see \Sec{Ino}) and depends only on $F_\phi$.  Meanwhile, the selectron mass 
depends on $f_1 \sim f_3 \equiv f$, which we can set equal to each other as an 
approximation, and $F_\phi$.  Given that the selectrons must be heavier then 
the \abbr{LSP}, for a viable dark matter candidate, yields
\begin{equation}
	f(F_\phi) \gtrsim 0.58.
\end{equation}

%|%%%%%%%%%%%%%%%%%%%%%%%%%%%%%%%DO NOT EXCEED%%%%%%%%%%%%%%%%%%%%%%%%%%%%%%%|%

%%%%%%%%%%%%%%%%%%%%%%%%%%%%%%%%%%%%%%%%%%%%%%%%%%%%%%%%%%%%%%%%%%%%%%%%%%%%%%%
% % % % % % % % % % % % % % % % % % % % % % % % % % % % % % % % % % % % % % % %
\subsection{Squarks}
% % % % % % % % % % % % % % % % % % % % % % % % % % % % % % % % % % % % % % % %
%%%%%%%%%%%%%%%%%%%%%%%%%%%%%%%%%%%%%%%%%%%%%%%%%%%%%%%%%%%%%%%%%%%%%%%%%%%%%%%

Squark masses in \abbr{mAMSB} decrease with energy due to the increase of 
$SU(2)_L$ and $U(1)_Y$ gauge couplings, which contribute negatively 
\cite{Paige:1999ui} to their masses.  At a certain energy scale, the negative 
contributions take over and the \abbr{AMSB} expressions for the squark soft 
masses become negative.  In our case this happens at an earlier scale due to 
the increase size of the $SU(2)_L$ and $U(1)_Y$ beta functions 
from the extra triplet and the doubly 
charged fields.  Normally we would have expected this to show up at high 
temperatures and lead to breakdown of color gauge symmetry. However, at 
high temperatures the vacuum of the theory is also affected by 
temperature corrections. Consequently the mass-square term of the squarks will 
have the form $\mu^2(T)_{\tilde{q}}\simeq (-M^2_{\text{AMSB}}+\lambda T^2)$.
 The first term only grows logarithmically with temperature whereas the second 
term grows quadratically. The coefficient $\lambda$ is positive so that the 
net effect is that $\mu^2(T)_{\tilde{q}}$ remains positive at high 
temperature and leaving color gauge symmetry intact in the 
early universe.

It is also worth noting that because non-asymptotically free 
gauge couplings contribute negatively to masses, the right-handed squarks are 
slightly heavier than left-handed squarks.  This is different than 
\abbr{mSUGRA} and \abbr{mGMSB} where all gauge couplings yield positive 
contributions making left-handed squarks heavier, see 
\fig{particle.spectrum.bosons}.  Furthermore, the squarks in this model can be 
degenerate with the sleptons.

%%%%%%%%%%%%%%%%%%%%%%%%%%%%%%%%%%%%%%%%%%%%%%%%%%%%%%%%%%%%%%%%%%%%%%%%%%%%%%%
% % % % % % % % % % % % % % % % % % % % % % % % % % % % % % % % % % % % % % % %
\subsection{Bosinos and the \abbr{LSP}}
\label{Sec:Ino}
% % % % % % % % % % % % % % % % % % % % % % % % % % % % % % % % % % % % % % % %
%%%%%%%%%%%%%%%%%%%%%%%%%%%%%%%%%%%%%%%%%%%%%%%%%%%%%%%%%%%%%%%%%%%%%%%%%%%%%%%

Because all superpartners eventually decay into the 
\abbr{LSP}, its makeup is an important part of \abbr{SUSY} collider 
phenomenology and dark matter prospects.  
Therefore understanding that makeup is an important task.  Cosmological 
constraints rule out a charged or colored \abbr{LSP} \cite{Ellis:1983ew}, hence 
limiting the choices to the sneutrino or the lightest neutralino.  
The former, in typical models, makes a 
poor dark matter candidate (relic 
abundances are too light; much 
of its mass range ruled out by direct detection 
\cite{Caldwell:1988su, Arina:2007tm}.  It is therefore more interesting 
to consider the lightest neutralino as the \abbr{LSP}, the candidate in 
common \abbr{SUSY} scenarios (except in \abbr{mGMSB} where it is the next 
to lightest \abbr{SUSY} particle but has the same collider significance
\cite{Giudice:1998bp}).  

%|%%%%%%%%%%%%%%%%%%%%%%%%%%%%%%%DO NOT EXCEED%%%%%%%%%%%%%%%%%%%%%%%%%%%%%%%|%

The lightest neutralino will be some mixture of the wino, bino and Higgsino.  
Its gaugino composition follows from the gaugino mass ratio 
which is easily calculated and relatively independent of the point in 
parameter space.  In \abbr{AMSB} this ratio depends on both the gauge 
couplings and the gauge coupling beta functions, $b$.  
The latter is important since this is where the effects of the light 
triplets and doubly-charged Higgs are felt (see \tbl{VALUES.b} for $b$ for values 
in \thismodel{} compare to \abbr{AMSB} based on \abbr{MSSM} particle content).  
It is calculated to be: $M_3:M_2:M_1 \sim 1.3:1:1.3$.  The striking 
characteristic of this ratio is its degeneracy when compared to 
\abbr{mSUGRA}/\abbr{mGMSB}, $M_3:M_2:M_1 \sim 3:1:0.3$ or even 
\abbr{mAMSB} $M_3:M_2:M_1 \sim 8:1:3.5$.

Specifically then, the \abbr{LSP} will have a 
large wino component where in \abbr{mAMSB} it is all wino, and there will also be 
some non-negligible mixing with the bino.  Note that in \abbr{mSUGRA} 
(\abbr{mGMSB}) the sole contribution to this ratio is from the gauge 
couplings and therefore the \abbr{LSP} (\abbr{NLSP}) is always mostly bino.  The Higgsino 
contribution is not independent of other parameters and therefore 
is not as predictable, but numerical results show 
that it is typically a little bit lighter than the wino (its value decreases 
compared to the wino as $F_\phi$ is increased).  Therefore, the 
\abbr{LSP} will be some combination mostly Higgsino with significant wino 
content and  and a little bit of bino.  The mixed Higgsino state will 
correspond to $\chi_2^0$; $\chi_3^0$ and $\chi_2^+$ will be mostly wino with 
some Higgsino (percent values will be complementary to those of $\chi_1$), and 
$\chi_4^0$ will be mostly bino.

%|%%%%%%%%%%%%%%%%%%%%%%%%%%%%%%%DO NOT EXCEED%%%%%%%%%%%%%%%%%%%%%%%%%%%%%%%|%

\begin{table}[ht]
\begin{center}
\begin{tabular}{c|c|c|c}
			&	$\quad{} b_1 \quad{}$
			&	$\quad{} b_2 \quad{}$
			&	$\quad{} b_3 \quad{}$
			\\
\hline
			\abbr{MSSM}
			&	$\frac{33}{5}$ 		%mSUGRA and mGMSB
			&	$1$			%mAMSB
			&	$-3$		%This model
			\\
			$\thismodel$	
			&	$\frac{78}{5}$ 		%mSUGRA and mGMSB
			&	$6$			%mAMSB
			&	$-3$		%This model			
\end{tabular}
\end{center}

\caption{
Values for the b parameter in the \abbr{MSSM} and \thismodel.  Note the 
larger values in \thismodel{} for $SU(2)_L$ and $U(1)_Y$.
}

\label{Table:VALUES.b}
\end{table}

%|%%%%%%%%%%%%%%%%%%%%%%%%%%%%%%%DO NOT EXCEED%%%%%%%%%%%%%%%%%%%%%%%%%%%%%%%|%

An immediate consequence of the degeneracy of the gauginos is a more natural 
heavy \abbr{LSP}, closer in mass to both the gluinos and squarks.  
Naturalness suggests that squark masses are not much larger than $1$ TeV, to 
minimize fine tuning in the Higgs mass, therefore:
\begin{align*}
	F_\phi	&	\lesssim 63 \text{ TeV}
\end{align*}
yielding:
\begin{align}
	M_1			&	\lesssim 1350 \text{ GeV}
	\notag
	\\
	M_2			&	\lesssim 980 \text{ GeV}
	\label{Eq:Naturalness.Gaugino.Masses}
\end{align}
This is a much larger value than the upper bound in \abbr{mAMSB} 
$M_2 \lesssim 200$ GeV \cite{Feng:1999hg} and therefore has less of its 
parameter spaced ruled out by experimental data.

%|%%%%%%%%%%%%%%%%%%%%%%%%%%%%%%%DO NOT EXCEED%%%%%%%%%%%%%%%%%%%%%%%%%%%%%%%|%

Another point to consider is that the Higgsino and wino form isospin 
doublets and triplets with the 
appropriate charginos.  Therefore when they play the role of the lightest 
neutralino, there is potential for a very small mass difference between the 
lightest neutralino and the lightest chargino.  This is very pronounced 
in \abbr{mAMSB} where the mass difference of the mostly wino neutralino and 
chargino is on the order of $100$s of MeVs including leading radiative 
corrections.  Analytical approximations for this quantity for large $\mu$ have 
been given in \cite{Gherghetta:1999sw, Feng:1999fu, 
Feng:1999hg}.  Such approximations are not as useful in \thismodel{} 
since the relevant mass scales: $\mu$, $M_1$, and $M_2$ are relatively of the 
same order (the singlino contribution is much larger than these); however, an 
analytic expression for the minimum of the mass difference is attainable.

%|%%%%%%%%%%%%%%%%%%%%%%%%%%%%%%%DO NOT EXCEED%%%%%%%%%%%%%%%%%%%%%%%%%%%%%%%|%

First note that a Higgsino mixing exists in the neutralino matrix, absent in 
the chargino sector.  This mixing goes to zero as $\tan{\beta} \rightarrow 1$ 
hence indicating that for $\tan{\beta} = 1$ the mass difference is minimal 
(when $\tan{\beta} \rightarrow 1$ and $\tan {\theta_W} \rightarrow 0$ the 
global custodial $SU(2)$ becomes an exact symmetry making the mass difference 
zero).  The eigenvalues of the two matrices can than be expanded for 
$\tan{\beta} = 1$ 
using the approximation $M_1 \sim M_2 > \mu \gg \MZ$, this yields, 
to first order:
\begin{equation}
\Delta_{\tilde \chi_1}
	\equiv 	m_{\tilde \chi_1^{\pm}} - m_{\tilde \chi_1^{0}} 
	>	2 \sin^2{\theta_W} \frac{\MZ^2}{M_1}
	\label{Eq:Ino.Mass.Difference}
\end{equation}

%|%%%%%%%%%%%%%%%%%%%%%%%%%%%%%%%DO NOT EXCEED%%%%%%%%%%%%%%%%%%%%%%%%%%%%%%%|%

The second order term is positive definite so that 
$\Delta_{\tilde \chi_1}$ can in fact be used as a minimal value for the mass 
splitting.  Notice that the $\Delta_{\tilde \chi_1} \rightarrow 0$ as 
$\tan{\theta_W} \rightarrow 0$ as argued above 
(and that $\Delta_{\tilde \chi_1} \rightarrow 0$ as 
$M_1 \rightarrow \infty$ since this also restores the custodial symmetry 
when $\tan{\beta} = 1$.  

%|%%%%%%%%%%%%%%%%%%%%%%%%%%%%%%%DO NOT EXCEED%%%%%%%%%%%%%%%%%%%%%%%%%%%%%%%|%

\begin{figure}[ht]

\begin{center}
	\begin{picture}(288,180)
		\put(0,0){\includegraphics[scale=1]{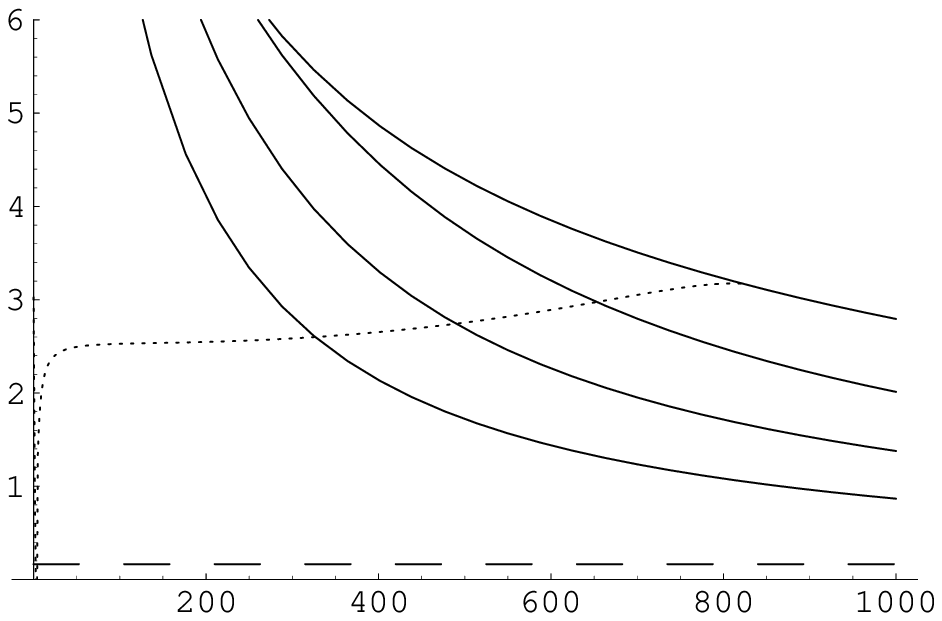}}
		\put(144, -10){$\mu$ (GeV)}
		\put(-20, 90){\parbox{1cm}{$\Delta_{\tilde \chi_1}$ (GeV)}}
		\put(268, 90){\scriptsize{$M_2 = 1.1 \mu$}}
		\put(268, 69){\scriptsize{$M_2 = 1.5 \mu$}}
		\put(268, 52){\scriptsize{$M_2 = 2 \mu$}}
		\put(268, 37){\scriptsize{$M_2 = 3 \mu$}}
		\put(30, 20){\scriptsize{\abbr{mAMSB}}}
		\put(30, 85){\scriptsize{$\MSUSY \sim 1$ TeV}}
	\end{picture}	
\end{center}

\caption{
	Mass difference of the lightest chargino and neutralino as a function 
	of 
	$\mu$ for $\lambda = 0.26, \tan{\beta} = 3.25$ and the singlino mass 
	term 
	$2 \inp{\mu_{\Slight} + \frac{1}{\sqrt{2}} \Slightvev \kappa} = 2 M_1$.
	From top to 
	bottom, $M_2 = 1.1 \mu$, $1.5 \mu$, $2 \mu$ and $3 \mu$.  The line at 
	$0.165$ GeV is the asymptotic value for large $M_2$ in \abbr{mAMSB}, while 
	the dotted curve is represents where squark masses are at 
	about a TeV.  Below this curve, the Higgs mass is somewhat fine-tuned.
}
\label{Fig:Ino.Mass.Difference}
\end{figure}

The form of \eq{Ino.Mass.Difference} is convenient since the only 
free parameter it depends on is $F_\phi$ (through $M_1$), which also 
controls the squark masses.  Applying the natural upper bound for $M_1$ from 
\eq{Naturalness.Gaugino.Masses} yields:
\begin{align}
	\Delta_{\tilde \chi_1}	&	> 1.4 \text{ GeV}
	\label{Eq:Ino.Mass.Difference.Minimum}
\end{align}
This is larger than the \abbr{mAMSB} value of a few $100$s of MeV.  Exact values for 
the mass difference are given in 
\fig{Ino.Mass.Difference} as a function of 
$\mu \equiv \frac{1}{\sqrt{2}} \lambda \Slightvev$ with: 
$\lambda = 0.26$, $\tan{\beta} = 3.25$ and the singlino mass 
term $2 \inp{\mu_{\Slight} + \frac{1}{\sqrt{2}} \Slightvev \kappa} = 2 M_1$.  
The line at $165$ GeV represents the asymptotic value for large $M_2$ in 
\abbr{mAMSB} at the one loop level\cite{Gherghetta:1999sw} and below the 
dotted line the squark masses are above a TeV and hence the Higgs mass becomes 
fined tuned.

%|%%%%%%%%%%%%%%%%%%%%%%%%%%%%%%%DO NOT EXCEED%%%%%%%%%%%%%%%%%%%%%%%%%%%%%%%|%

%|%%%%%%%%%%%%%%%%%%%%%%%%%%%%%%%DO NOT EXCEED%%%%%%%%%%%%%%%%%%%%%%%%%%%%%%%|%

%%%%%%%%%%%%%%%%%%%%%%%%%%%%%%%%%%%%%%%%%%%%%%%%%%%%%%%%%%%%%%%%%%%%%%%%%%%%%%%
% % % % % % % % % % % % % % % % % % % % % % % % % % % % % % % % % % % % % % % %
\subsection{Collider Signatures}
\label{Sec:Collider}
% % % % % % % % % % % % % % % % % % % % % % % % % % % % % % % % % % % % % % % %
%%%%%%%%%%%%%%%%%%%%%%%%%%%%%%%%%%%%%%%%%%%%%%%%%%%%%%%%%%%%%%%%%%%%%%%%%%%%%%%

The small size of $\Delta_{\tilde \chi_1}$ can potentially be problematic at 
a collider because the soft decay products, $X$, in the process $\chi_1^+ \rightarrow 
X \chi_1^0$, will not be visible.  This is a feature shared by both 
\thismodel{} and \abbr{mAMSB}.  The difference is that the larger value of 
$\Delta_{\tilde \chi_1}$ for \thismodel{} might produce prospects 
of detection if $X = \tau$ or a hard $\mu$; however, this advantage is 
counterbalanced by a faster chargino decay eliminating chances of long-lived 
charged tracks with no muon chamber activity.  Regardless, similar situations have 
been analyzed and found to be manageable for both letpon colliders\cite{Chen:1995yu} 
and the Tevatron \cite{Feng:1999fu, Gunion:1999jr}.

On the other hand, \abbr{LHC} studies of 
\abbr{mAMSB} have focused on \abbr{mSUGRA} like signals 
\cite{Baer:2000bs, Paige:1999ui}.  Such signals are heavily dependent on 
lepton final states and are based on left-handed squark decays to mostly 
wino charginos and neutralinos.  These in turn can decay leptonically 
producing trilepton signals or same sign dilepton signals 
\cite{Hinchliffe:1996iu, Martin:1997ns}, both of which have potentially 
manageable backgrounds.  For in \abbr{mAMSB}, though, the 
the wino states are the lightest and will not decay leptonically.  Hence the 
the right-handed squarks take the place of the left-handed ones decaying into 
the mostly bino neutralino (which can decay leptonically).  Yet, since there is 
no corresponding chargino to the bino, signals such as 
the trilepton and the same sign dilepton signal may not be possible.

The situation in \thismodel{} is more analogous to \abbr{mSUGRA}: right-handed 
squarks will decay to the \abbr{LSP} which has some bino content.  Meanwhile, the 
left-handed squarks may decay either to the lightest chargino/neutralino, or, 
more likely (because of their higher wino content), to $\chi_3^0$ and $\chi_2^+$.  
These may then decay leptonically depending on the slepton masses 
(\textit{e.g} $f(v_R) = 1.4$ in \tbl{Spectrum}) giving the 
familiar signals: trilepton and same sign dilepton.  Note that it 
is also possible that decay of $\chi_1^+$ will produce leptonic signals since 
$\Delta_{\tilde \chi_1}$ is larger.  These considerations would help differntiate 
this model from \abbr{mAMSB}, while the degeneracies in the gaugino sector 
and same generation slepton will differentiate it from 
\abbr{mSUGRA} and \abbr{mGMSB}.  These differences between the various
scenarios are summarized in 
\tbl{Differences}.

%|%%%%%%%%%%%%%%%%%%%%%%%%%%%%%%%DO NOT EXCEED%%%%%%%%%%%%%%%%%%%%%%%%%%%%%%%|%

\begin{table}[ht]
\begin{center}
\begin{tabular}{c|c|c|c}
			& \abbr{mSUGRA}
			&
			&
			\\
			&	 \; and \abbr{mGMSB} \;
			&	\; \abbr{mAMSB} \;
			&	\; \thismodel
			\\
\hline
$M_3:M_2:M_1$	
			&	$3:1:0.3$ 		%mSUGRA and mGMSB
			&	$8:1:3.5$			%mAMSB
			&	$1.3:1:1.3$		%This model
			\\
$\abs{M_1},\abs{M_2}$ (GeV) Naturalness upperbound
			&	$130,260$\footnote{\abbr{mGMSB} only, in \abbr{mSUGRA} sfermion and gaugino mass are 
determined by two seperate parameters.} 		%mSUGRA and mGMSB
			&	$640, 200$				%mAMSB
			&	$1350, 980$			%This model
			\\
Same generation slepton mass percent difference
			&	$\sim 150 \%$ 		%mSUGRA and mGMSB
			&	$\sim 2\%$				%mAMSB
			&	$\sim 4\%$			%This model
			\\
Possibility of slepton-squark degeneracy
			&	No 		%mSUGRA and mGMSB
			&	No		%mAMSB
			&	Yes		%This model
\end{tabular}
\end{center}

\caption
{
	A list of phenomenological characteristics of interest in \abbr{mSUGRA}, 
	\abbr{mGMSB}, \abbr{mAMSB} and \thismodel.
}

\label{Table:Differences}
\end{table}

%|%%%%%%%%%%%%%%%%%%%%%%%%%%%%%%%DO NOT EXCEED%%%%%%%%%%%%%%%%%%%%%%%%%%%%%%%|%

%%%%%%%%%%%%%%%%%%%%%%%%%%%%%%%%%%%%%%%%%%%%%%%%%%%%%%%%%%%%%%%%%%%%%%%%%%%%%%%
% % % % % % % % % % % % % % % % % % % % % % % % % % % % % % % % % % % % % % % %
\subsection{Triplets and Doubly-Charged Higgses}
\label{Sec:DC}
% % % % % % % % % % % % % % % % % % % % % % % % % % % % % % % % % % % % % % % %
%%%%%%%%%%%%%%%%%%%%%%%%%%%%%%%%%%%%%%%%%%%%%%%%%%%%%%%%%%%%%%%%%%%%%%%%%%%%%%%

The interplay between \abbr{AMSB} and the left-handed and doubly-charged Higgses leads 
to interesting phenomenology and is worth 
summarizing here.  Because they play the central role of saving the slepton masses 
from a tachyonic fate, their masses must be around the $F_\phi$ scale.  This
puts a bound on the right-handed scale scale, $v_R \lesssim 10^{12}$ GeV, which is not the 
case when these particles appear in \abbr{mSUGRA} and \abbr{mGMSB} 
\cite{Dutta:1998bn, Setzer:2006sf}.  It is also possible, through mixing due to 
bilinear $b$-terms \eq{DC.Mass}, that one triplet and one doubly charged Higgs will 
be light, $\mathcal{O}(1 \text{ TeV})$ and therefore accessible at the \abbr{LHC}.  
Their presence would also be felt indirectly in upcoming muonium-antimuonium oscillation
experiments 
since their couplings to first and second generation leptons must be large.  
For sleptons above the LEP II bound:
\begin{equation}
	f_1(F_\phi) \sim f_2(F_\phi) \sim f_{c1}(F_\phi) \sim f_{c2}(F_\phi) \sim 0.5
\end{equation}
and for sleptons above the lightest neutralino (for a good dark matter candidate):
\begin{equation}
	f_1(F_\phi) \sim f_2(F_\phi) \sim f_{c1}(F_\phi) \sim f_{c2}(F_\phi) \sim 0.6
\end{equation}
Based on \fig{f.FIXED.POINT.f3.Zero} and \fig{SLEPTON.MASS.CONTOUR}.
On the 
other hand, all the triplet and doubly-charged Higgsinos will remain heavy, 
$\mathcal{O}(F_\phi)$, and undetectable at the \abbr{LHC} or low energy experiments.

%|%%%%%%%%%%%%%%%%%%%%%%%%%%%%%%%DO NOT EXCEED%%%%%%%%%%%%%%%%%%%%%%%%%%%%%%%|%

%%%%%%%%%%%%%%%%%%%%%%%%%%%%%%%%%%%%%%%%%%%%%%%%%%%%%%%%%%%%%%%%%%%%%%%%%%%%%%%
% % % % % % % % % % % % % % % % % % % % % % % % % % % % % % % % % % % % % % % %
\subsection{Dark matter}
\label{DM}
% % % % % % % % % % % % % % % % % % % % % % % % % % % % % % % % % % % % % % % %
%%%%%%%%%%%%%%%%%%%%%%%%%%%%%%%%%%%%%%%%%%%%%%%%%%%%%%%%%%%%%%%%%%%%%%%%%%%%%%%

As noted in the previous section, the \abbr{LSP} in our model is a predominatly 
Higgsino wino mix with very little bino (about $1 \%$). 
Since the annihilation rate for such an \abbr{LSP} is large, its 
relic density from conventional annihilation arguments is not enough to 
explain the observed $\Omega_m$ of the universe of 20\%. This issue has 
been discussed earlier in ref.\cite{Moroi:1999zb}, according to which the 
decay of the gravitino in the late stage of the universe to non-thermal 
winos will generate enough density to make it a viable dark matter. A similar 
mechanism would work in this mostly Higgsino case since the crucial ingredients 
are similar: the \abbr{LSP} mass (this a similar), its interactions with the 
gravitino (again, this is similar between the two cases because of the similar 
masses) and its annihilation rate 
(these are also the same since wino and Higgsino annihilation takes place through 
a t-channel chargino exchange proportional with $\alpha_2$ strength).  
Like \cite{Moroi:1999zb}, we have scanned over the 
parameters and found that such dark matter does evade current bounds on 
direct detection set by CDMS Soudan and EDELWEISS but will be detectable by 
future experiments.

%|%%%%%%%%%%%%%%%%%%%%%%%%%%%%%%%DO NOT EXCEED%%%%%%%%%%%%%%%%%%%%%%%%%%%%%%%|%

%%%%%%%%%%%%%%%%%%%%%%%%%%%%%%%%%%%%%%%%%%%%%%%%%%%%%%%%%%%%%%%%%%%%%%%%%%%%%%%
%%%%%%%%%%%%%%%%%%%%%%%%%%%%%%%%%%%%%%%%%%%%%%%%%%%%%%%%%%%%%%%%%%%%%%%%%%%%%%%
\section{Beyond $v_R$}
\label{Sec:Miscelleneous}
%%%%%%%%%%%%%%%%%%%%%%%%%%%%%%%%%%%%%%%%%%%%%%%%%%%%%%%%%%%%%%%%%%%%%%%%%%%%%%%
%%%%%%%%%%%%%%%%%%%%%%%%%%%%%%%%%%%%%%%%%%%%%%%%%%%%%%%%%%%%%%%%%%%%%%%%%%%%%%%

In this section, we comment on the ultraviolet behaviour of the theory. 
 As we see from Figure \ref{Fig:Gauge.Couplings} below, despite the 
new contributions to $SU(2)_L$ and $U(1)_Y$ beta functions, all 
couplings remain perturbative until about $10^{11}$--$10^{12}$ GeV. Our 
effective field theory approach below this scale should hold without any 
problem. Once we are above this scale, the couplings could maintain 
perturbativity if there are extra dimensions\cite{Dienes:1998vh} due to 
negative contributions from vector gauge KK modes of the theory if the 
inverse radius of the extra dimensions are around $10^{11}$ GeV or so. 
Such extra dimensions could also be the origin of the Planck suppressed 
operators that we have used in our discussion.

\begin{figure}[ht]
	\begin{center}
		\begin{picture}(288,180)
			\put(0,0){\includegraphics[scale=1]{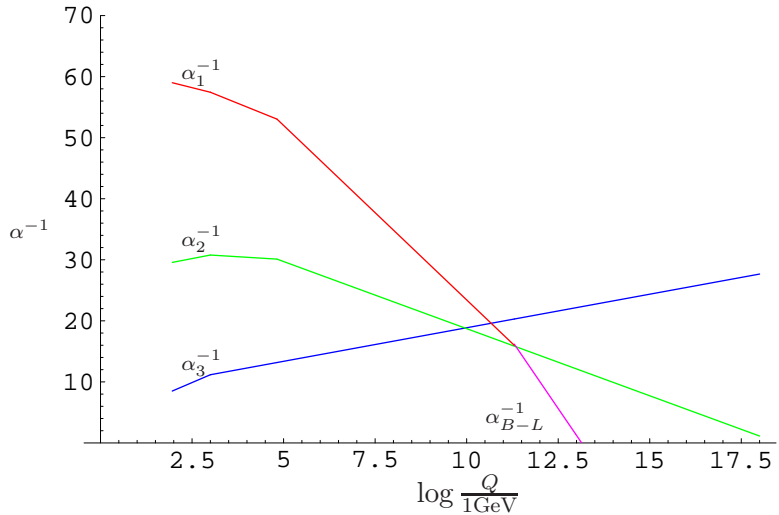}}
			\put(144, -10){$\log{\frac{Q}{1 \text{GeV}}}$}
			\put(-10, 90){\scriptsize{$\alpha^{-1}$}}
			\put(55, 150){\scriptsize{$\alpha_1^{-1}$}}
			\put(55, 87){\scriptsize{$\alpha_2^{-1}$}}
			\put(55, 40){\scriptsize{$\alpha_3^{-1}$}}
			\put(170, 20){\scriptsize{$\alpha_{B-L}^{-1}$}}
		\end{picture}

	\end{center}

	\caption{
		Inverse gauge couplings as a function of the $\log$ of the energy scale.  
		The $v_R$ scale 
		is at about $10^{12}$ GeV at which point $\alpha_R^{-1}$ begins to run with 
		its curve being indistinguishable from $\alpha_2^{-1}$ due to parity.  
	}
	\label{Fig:Gauge.Couplings}
\end{figure}

%|%%%%%%%%%%%%%%%%%%%%%%%%%%%%%%%DO NOT EXCEED%%%%%%%%%%%%%%%%%%%%%%%%%%%%%%%|%

%%%%%%%%%%%%%%%%%%%%%%%%%%%%%%%%%%%%%%%%%%%%%%%%%%%%%%%%%%%%%%%%%%%%%%%%%%%%%%%
%%%%%%%%%%%%%%%%%%%%%%%%%%%%%%%%%%%%%%%%%%%%%%%%%%%%%%%%%%%%%%%%%%%%%%%%%%%%%%%
\section{Conclusion}
%%%%%%%%%%%%%%%%%%%%%%%%%%%%%%%%%%%%%%%%%%%%%%%%%%%%%%%%%%%%%%%%%%%%%%%%%%%%%%%
%%%%%%%%%%%%%%%%%%%%%%%%%%%%%%%%%%%%%%%%%%%%%%%%%%%%%%%%%%%%%%%%%%%%%%%%%%%%%%%

In summary, we have elaborated on our suggestion that minimally extending 
\abbr{MSSM} to account for neutrino masses in a way that $R$-parity remains an 
automatic symmetry of the theory allows for a solution to the tachyonic 
slepton problem of anomaly mediated supersymmetry breaking. 
Interestingly, the solution requires that parity symmetry remain exact 
above the $v_R$ scale. Among the new results, we show how 
to obtain radiative electroweak symmetry breaking and a reasonable $B\mu$ term
in this class of models. We also discuss the sparticle spectrum of the 
model in detail and show how it differs from that of \abbr{mAMSB} 
as well as other widely discussed 
supersymmetry scenarios. A new feature of this model is the presence of 
new TeV scale $SU(2)_L$ triplets and doubly charged $SU(2)_L$ singlet 
fields, whose phenomenology has been the subject of many 
papers
\cite{Dine:2007xi,Han:2007bk,Frank:2007nv,Raidal:1998vi,Barenboim:1996pt, 
Akeroyd:2007zv}. We believe that the model 
discussed here is a serious alternative to the \abbr{mAMSB} whose further 
phenomenological implications need to be explored in detail.

%%%%%%%%%%%%%%%%%%%%%%%%%%%%%%%%%%%%%%%%%%%%%%%%%%%%%%%%%%%%%%%%%%%%%%%%%%%%%%%
%%%%%%%%%%%%%%%%%%%%%%%%%%%%%%%%%%%%%%%%%%%%%%%%%%%%%%%%%%%%%%%%%%%%%%%%%%%%%%%
\section{Acknowledgments}
\label{Sec:Acknowledgments}
%%%%%%%%%%%%%%%%%%%%%%%%%%%%%%%%%%%%%%%%%%%%%%%%%%%%%%%%%%%%%%%%%%%%%%%%%%%%%%%
%%%%%%%%%%%%%%%%%%%%%%%%%%%%%%%%%%%%%%%%%%%%%%%%%%%%%%%%%%%%%%%%%%%%%%%%%%%%%%%

%|%%%%%%%%%%%%%%%%%%%%%%%%%%%%%%%DO NOT EXCEED%%%%%%%%%%%%%%%%%%%%%%%%%%%%%%%|%

This work is supported by the National Science Foundation
Grant No. PHY-0652363. We like to thank Z. Chacko for many useful 
comments and B. Dutta for discussions.

%%%%%%%%%%%%%%%%%%%%%%%%%%%%%%%%%%%%%%%%%%%%%%%%%%%%%%%%%%%%%%%%%%%%%%%%%%%%%%%
%   |   %   |   %   |   %   |   %   |   %   |   %   |   %   |   %   |   %   | %
\appendix
%   |   %   |   %   |   %   |   %   |   %   |   %   |   %   |   %   |   %   | %
%%%%%%%%%%%%%%%%%%%%%%%%%%%%%%%%%%%%%%%%%%%%%%%%%%%%%%%%%%%%%%%%%%%%%%%%%%%%%%%

%%%%%%%%%%%%%%%%%%%%%%%%%%%%%%%%%%%%%%%%%%%%%%%%%%%%%%%%%%%%%%%%%%%%%%%%%%%%%%%
%%%%%%%%%%%%%%%%%%%%%%%%%%%%%%%%%%%%%%%%%%%%%%%%%%%%%%%%%%%%%%%%%%%%%%%%%%%%%%%
\section{Notation Conventions}
\label{App:Notation.Conventions}
%%%%%%%%%%%%%%%%%%%%%%%%%%%%%%%%%%%%%%%%%%%%%%%%%%%%%%%%%%%%%%%%%%%%%%%%%%%%%%%
%%%%%%%%%%%%%%%%%%%%%%%%%%%%%%%%%%%%%%%%%%%%%%%%%%%%%%%%%%%%%%%%%%%%%%%%%%%%%%%

In this appendix we summarize our notational conventions.  
Given a superpotential defined as
\begin{equation}
W =
  L^i \Phi_i 
+ \frac{1}{2!} \mu^{ij} \Phi_i \Phi_j
+ \frac{1}{3!} Y^{ijk} \Phi_i \Phi_j \Phi_k
+ \frac{1}{4!} \frac{\lambda^{ijk\ell}}{M} \Phi^i \Phi^j \Phi^k \Phi^\ell
+ \cdots
\end{equation}
with a corresponding lagrangian of
\begin{equation}
\mathcal{L} =
  \intOp[4]{\theta}\inp{Z^i_j \Phi_i \Phi^{j *} + \cdots}
+ \inb{
	  \intOp[2]{\theta} 
	  	\inp{
			 W
			+ \mathscr{W}^\alpha \mathscr{W}_\alpha 
		}
	+ \text{h.c.}
	}
\end{equation}
the anomalous dimensions, $\gamma^i_j$, and $\beta$-functions, 
$\beta_{L}^{i}$, $\beta_{\mu}^{ij}$, $\beta_{Y}^{ijk}$, 
at a given energy scale $Q$ are defined by
\begin{align}
\gamma^i_j
	& = \deriv{\ln Z^i_j}{\ln Q}
	  = 4 C_a\inp{\Phi_i} g_a^2 \delta^i_j - Y_{jpq} Y^{ipq}
	\\
\beta_L^{i}
	& =	\deriv{L^i}{\ln Q}
	  =	- \half L^{j} \gamma^i_j 
	\\
\beta_\mu^{ij}
	& =	\deriv{\mu^{ij}}{\ln Q}
	  =	- \half \mu^{ip} \gamma^j_p 
		+ \inp{j \leftrightarrow i}
	\\
\beta_Y^{ijk}
	& =	\deriv{Y^{ijk}}{\ln Q}
	  =	- \half Y^{ijp} \gamma^j_p 
		+ \inp{k \leftrightarrow i}
		+ \inp{k \leftrightarrow j}
\end{align}

Furthermore, we choose the sign of the soft \abbr{SUSY} breaking terms by 
specifying that
\begin{equation}
V_{\text{SB}} =
  \half \inp{m^2}^i_j \scalar{\Phi}_i \scalar{\Phi}^{j *}
+ \ell^i \scalar{\Phi}_i
+ \frac{1}{2!} b^{ij} \scalar{\Phi}_i \scalar{\Phi}_j
+ \frac{1}{3!} a^{ijk} \scalar{\Phi}_i \scalar{\Phi}_j \scalar{\Phi}_k
+ \text{h.c.}
\end{equation}

%%%%%%%%%%%%%%%%%%%%%%%%%%%%%%%%%%%%%%%%%%%%%%%%%%%%%%%%%%%%%%%%%%%%%%%%%%%%%%%
%%%%%%%%%%%%%%%%%%%%%%%%%%%%%%%%%%%%%%%%%%%%%%%%%%%%%%%%%%%%%%%%%%%%%%%%%%%%%%%
\section{Between scales: $v_R$ To $F_\phi$}
\label{App:Between.Scales}
%%%%%%%%%%%%%%%%%%%%%%%%%%%%%%%%%%%%%%%%%%%%%%%%%%%%%%%%%%%%%%%%%%%%%%%%%%%%%%%
%%%%%%%%%%%%%%%%%%%%%%%%%%%%%%%%%%%%%%%%%%%%%%%%%%%%%%%%%%%%%%%%%%%%%%%%%%%%%%%

The superpotential between the $v_R$ and $F_\phi$ scale is:
\begin{align}
\notag
W_{\text{NMSSM}++}
	& =	\I y_t^a Q^T \tau_2 H_{ua} t^c + \I y_b^a Q^T \tau_2 H_{da} b^c
								+ \I y_\tau^a Q^T \tau_2 H_{da} t^c
								\\
								\notag
								&
								\quad{} + f_{ci} e^c_i \Delta^{c--} e^c_i
								+ \I f_{i} L_i^T \tau_2 \Delta L_i
								\\
								&
								\quad
								+ \I \lambda^{ab} \Slight H_{ua} \tau_2 H_{db}
								+ \half \mu_\Slight \Slight^2 + \third \kappa \Slight^3
\end{align}
Where $a=1,2$, the \abbr{MSSM} Yukawa matrices have been approximated by 
the third generation diagonal term, the seesaw Yukawa couplings are 
diagonal and the subscript $i$ represents lepton generation and is summed.  
The gamma functions for the theory between the $v_R$ and $F_\phi$ are:

\begin{align}
	\gamma_{Q_3}	&	=	- \frac{1}{8 \pi^2}
										\inp
										{
											y_t^{a*} y_t^{a} + y_b^{a*} y_b^{a} -
											\frac{8}{3} g_3^2 - \frac{3}{2} g_2^2 - 
											\frac{1}{30} g_1^2
										}
	\\
	\gamma_{Q_1}	&	=	- \frac{1}{8 \pi^2}
										\inp
										{
											-\frac{8}{3} g_3^2 - \frac{3}{2} g_2^2 - 
											\frac{1}{30} g_1^2
										}
	\\
	\gamma_{t^c}	&	=	- \frac{1}{8 \pi^2}
										\inp
										{
											2 y_t^{a*} y_t^{a} -
											\frac{8}{3} g_3^2 - \frac{8}{15} g_1^2
										}
	\\
	\gamma_{u^c}	&	=	- \frac{1}{8 \pi^2}
										\inp
										{
											-\frac{8}{3} g_3^2 - \frac{8}{15} g_1^2
										}
	\\
	\gamma_{b^c}	&	=	- \frac{1}{8 \pi^2}
										\inp
										{
											2 y_b^{a*} y_b^{a} -
											\frac{8}{3} g_3^2 - \frac{2}{15} g_1^2
										}
	\\
	\gamma_{d^c}	&	=	- \frac{1}{8 \pi^2}
										\inp
										{
											-\frac{8}{3} g_3^2 - \frac{2}{15} g_1^2
										}
	\\
	\gamma_{L_3}	&	=	- \frac{1}{8 \pi^2}
										\inp
										{
											y_{\tau}^{a*} y_{\tau}^{a} + 6 \abs{f_3}^2
											- \frac{3}{2} g_2^2 - \frac{3}{10} g_1^2
										}
	\\
	\gamma_{L_1}	&	=	- \frac{1}{8 \pi^2}
										\inp
										{
											6 \abs{f_1}^2
											- \frac{3}{2} g_2^2 - \frac{3}{10} g_1^2
										}
	\\
	\gamma_{\tau^c}	&	=	- \frac{1}{8 \pi^2}
										\inp
										{
											2 y_{\tau}^{a*} y_{\tau}^{a} + 4 \abs{f_{c3}}^2
											- \frac{6}{5} g_1^2
										}
	\\
	\gamma_{e^c}	&	=	- \frac{1}{8 \pi^2}
										\inp
										{
											4 \abs{f_{c1}}^2
											- \frac{6}{5} g_1^2
										}
\end{align}
\begin{align}
	\gamma_{\Slight}	&	=	- \frac{1}{8 \pi^2}
										\inp
										{
											2 \abs{\kappa}^2 + 2 \lambda^{ab*} \lambda^{ab}
										}
	\\
	\gamma_{H_{ua}}^{H_{ub}}	&	=	- \frac{1}{8 \pi^2}
										\inp
										{
											3 y_t^{a*} y_t^{b} + \lambda^{ac} \lambda^{bc}
											- \delta^{ab}
											\inp
											{
												\frac{3}{2} g_2^2 + \frac{3}{10} g_1^2
											}
										}
	\\
	\gamma_{H_{da}}^{H_{db}}	&	=	- \frac{1}{8 \pi^2}
										\inp
										{
											3 y_b^{a*} y_b^{b} + y_\tau^{a*} y_\tau^{b}
											+ \lambda^{ca*} \lambda^{cb}
											- \delta^{ab}
											\inp
											{
												\frac{3}{2} g_2^2 + \frac{3}{10} g_1^2
											}
										}
	\\
	\gamma_{\Delta}	&	=	- \frac{1}{8 \pi^2}
										\inp
										{
											2 \abs{f_3}^2 + 2 \abs{f_2}^2 + 2 \abs{f_1}^2
											- 4 g_2^2 - \frac{6}{5} g_1^2
										}
	\\
	\gamma_{\bar{\Delta}}	&	=	- \frac{1}{8 \pi^2}
										\inp
										{
											- 4 g_2^2 - \frac{6}{5} g_1^2
										}
	\\
	\gamma_{\Delta^{c--}}	&	=	- \frac{1}{8 \pi^2}
										\inp
										{
											2 \abs{f_c3}^2 + 2 \abs{f_c2}^2 + 2 \abs{f_c1}^2
											- \frac{24}{5} g_1^2
										}
	\\
	\gamma_{\bar \Delta^{c--}}	&	=	- \frac{1}{8 \pi^2}
										\inp
										{
											- \frac{24}{5} g_1^2
										}
\end{align}

These expressions were used for the slepton masses in Eqs.~\eqn{AMSB.Mass.Selectron.Right} and
\eqn{AMSB.Mass.Selectron.Left}. The third generation squark masses can also be written down 
(here we assume real yukawa couplings for simplicity):
\begin{align}
	\notag
	m_{Q_3}^2
		&	=	
		\frac{1}{4} F_\phi^2
		\left\{
			- \frac{b_1 \alpha_1^2}{72 \pi^2} - \frac{3 b_2 \alpha_2^2}{8 \pi^2}
			-\frac{2 b_3 \alpha_3^2}{3 \pi^2}
		\right.
		\\
		\notag
		&
		\left.
			\quad{} + \frac{4 y^a_t}{16 \pi^2}
			\left[
				\frac{y^{a}_t}{16 \pi^2}
				\inp
				{
					3 \inp{y^{c}_t}^2 + \inp{y_b^{c}}^2
					- \frac{13}{9} g_1^2 - 3 g_2^2 - \frac{8}{3} g_3^2
				}
				+ \frac{y^c_t}{16 \pi^2}
				\inp
				{
					3 y^a_t y^c_t + \lambda^{ad} \lambda^{cd}
				}
			\right]
		\right.
		\\
		\notag
		&
		\left.
			\quad{}
			+ \frac{4 y^a_b}{16 \pi^2}
			\left[
				\frac{y^a_b}{16 \pi^2}
				\inp
				{
					3 \inp{y^c_b}^2 + \inp{y^c_t}^2 + \inp{y^c_\tau}^2
					- \frac{7}{9} g_1^2 - 3 g_2^2 - \frac{16}{3} g_3^2
				}
			\right.
		\right.
		\\
		&
		\left.
			\left.
				\quad{} \quad{}
				+ \frac{y^c_b}{16 \pi^2}
				\inp
				{
					3 y^a_b y^c_b + y^a_\tau y^c_\tau + \lambda^{da} \lambda^{dc}
				}
			\right]
		\right\}
	\\
	\notag
	m_{t^c}^2
		&	=	
		\frac{1}{4} F_\phi^2
		\left\{
			- \frac{2 b_1 \alpha_1^2}{9 \pi^2} -\frac{2 b_3 \alpha_3^2}{3 \pi^2}
		\right.
		\\
		&
		\left.
			\quad{} + \frac{8 y^a_t}{16 \pi^2}
			\left[
				\frac{y^{a}_t}{16 \pi^2}
				\inp
				{
					3 \inp{y^{c}_t}^2 + \inp{y_b^{c}}^2
					- \frac{13}{9} g_1^2 - 3 g_2^2 - \frac{8}{3} g_3^2
				}
				+ \frac{y^c_t}{16 \pi^2}
				\inp
				{
					3 y^a_t y^c_t + \lambda^{ad} \lambda^{cd}
				}
			\right]
		\right\}
		\\
		\notag
		m_{b^c}^2
		&	=	
		\frac{1}{4} F_\phi^2
		\left\{
			- \frac{b_1 \alpha_1^2}{18 \pi^2} - \frac{2 b_3 \alpha_3^2}{3 \pi^2}
		\right.
		\\
		\notag
		&
		\left.
			\quad{}
			+ \frac{8 y^a_b}{16 \pi^2}
			\left[
				\frac{y^a_b}{16 \pi^2}
				\inp
				{
					3 \inp{y^c_b}^2 + \inp{y^c_t}^2 + \inp{y^c_\tau}^2
					- \frac{7}{9} g_1^2 - 3 g_2^2 - \frac{16}{3} g_3^2
				}
			\right.
		\right.
		\\
		&
		\left.
			\left.
				\quad{} \quad{}
				+ \frac{y^c_b}{16 \pi^2}
				\inp
				{
					3 y^a_b y^c_b + y^a_\tau y^c_\tau + \lambda^{da} \lambda^{dc}
				}
			\right]
		\right\}
\end{align}

Were the first generation squark masses can be found by using the third generation 
mass expressions with yukawa couplings set to zero and $b_A = \inp{\frac{78}{5}, 6, -3}$ 
for $A = \inp{1,2,3}$. Typically, the largest contribution to these are given by:
\begin{eqnarray}
	m^2_{\tilde{q}}~\sim~F^2_\phi \frac{\alpha^2_3(F_\phi)}{2\pi^2}.
\end{eqnarray}

The present LEP bound on the Higgs mass of $114$ GeV can then
roughly be translated to give a lower bound of about $600$ GeV on the top 
squark 
mass. Using $\alpha_3(F_\phi)\simeq 0.08$ in the above expressions, we
can translate this squark mass bound to a lower limit on $F_\phi$ of 
about $30$ TeV. We have used this in all our calculations in the text.

\setcounter{section}{3}
\setcounter{equation}{0}

%%%%%%%%%%%%%%%%%%%%%%%%%%%%%%%%%%%%%%%%%%%%%%%%%%%%%%%%%%%%%%%%%%%%%%%%%%%%%%%
%%%%%%%%%%%%%%%%%%%%%%%%%%%%%%%%%%%%%%%%%%%%%%%%%%%%%%%%%%%%%%%%%%%%%%%%%%%%%%%
\section*{Addendum}
%%%%%%%%%%%%%%%%%%%%%%%%%%%%%%%%%%%%%%%%%%%%%%%%%%%%%%%%%%%%%%%%%%%%%%%%%%%%%%%
%%%%%%%%%%%%%%%%%%%%%%%%%%%%%%%%%%%%%%%%%%%%%%%%%%%%%%%%%%%%%%%%%%%%%%%%%%%%%%%

%|%%%%%%%%%%%%%%%%%%%%%%%%%%%%%%%DO NOT EXCEED%%%%%%%%%%%%%%%%%%%%%%%%%%%%%%%|%

The purpose of this addendum is to clarify certain aspects of the detailed 
model implementing the idea described in the main body of the paper. 
We first show that the model defined in 
Eqs.~\eqn{SuperW.SUSYLR}--\eqn{SuperW.SUSYLR.nr.gsv} has new diagrams at the 
$\Fphi$ scale that dominate the contributions noted in the text, making the 
sleptons tachyonic below $\Fphi$.  It is then noted that the model permits
an additional term in the k\"ahler potential that is crucial to restoring 
the low-energy phenomenology and leaves the presented results unaltered.

%|%%%%%%%%%%%%%%%%%%%%%%%%%%%%%%%DO NOT EXCEED%%%%%%%%%%%%%%%%%%%%%%%%%%%%%%%|%

To elucidate the issues at the $\Fphi$ scale, it is useful to first consider 
a simplified model with a superpotential of
\begin{equation}	
W_{\text{simp}}
	=	   \inp{\lambda_\Sheavy \Sheavy - M_\Delta \phi} 
			\inp{\DeltaC \DeltaBarC - M_{\Sheavy}^2 \phi^2}
		+ \frac{\lambda_A^c}{\MNP \phi} \Tr^2\inp{\DeltaC \DeltaBarC}
		+ \frac{\lambda_B^c}{\MNP \phi}\Tr\inp{\DeltaC \DeltaC}
			\Tr\inp{\DeltaBarC \DeltaBarC}
\label{Eq:Add.simple.superW.higgs}
\end{equation}
%
%|%%%%%%%%%%%%%%%%%%%%%%%%%%%%%%%DO NOT EXCEED%%%%%%%%%%%%%%%%%%%%%%%%%%%%%%%|%
%
and fields as defined in the text.  The mass scales $M_\Delta$, $M_{\Sheavy}$
are assumed to be of the same order as $v_R$, the right-handed scale.

%|%%%%%%%%%%%%%%%%%%%%%%%%%%%%%%%DO NOT EXCEED%%%%%%%%%%%%%%%%%%%%%%%%%%%%%%%|%
  
The superfields of \eq{Add.simple.superW.higgs} acquire a \abbr{VEV} given by
\begin{align}
\vev{\Sheavy}	& = \frac{M_\Delta}{\lambda_\Sheavy} \phi	\\
\vev{\DeltaC}	& = \vev{\DeltaBarC} = M_\Sheavy \phi			
\end{align}
%
%|%%%%%%%%%%%%%%%%%%%%%%%%%%%%%%%DO NOT EXCEED%%%%%%%%%%%%%%%%%%%%%%%%%%%%%%%|%
%
and, as expected, the \abbr{VEV}s are proportional to $\phi$ indicating this
is an \abbr{AMSB} preserving threshold.  It is worth noting that preserving
\abbr{AMSB} is a direct result of the superconformal invariance of the 
\abbr{VEV} structure which is itself a result of the \abbr{VEV}s being 
induced by terms that preserve the superconformal symmetry.

%|%%%%%%%%%%%%%%%%%%%%%%%%%%%%%%%DO NOT EXCEED%%%%%%%%%%%%%%%%%%%%%%%%%%%%%%%|%

Now once the superfields are shifted by their \abbr{VEV}s, the 
non-renormalizable terms give rise to an effective mass term for the 
(otherwise massless) doubly-charged fields:
\begin{equation}
W_{\text{simp}}
	\supset	\frac{M_\Sheavy^2 \phi^2}{\MNP \phi} \DCmm \DBarCpp 
	= \mu_{DC} \phi \DCmm \DBarCpp,
\label{Eq:Add.simple.superW.mass.DC}
\end{equation}
where $\mu_{DC} \equiv \frac{M_\Sheavy^2}{\MNP}$.

%|%%%%%%%%%%%%%%%%%%%%%%%%%%%%%%%DO NOT EXCEED%%%%%%%%%%%%%%%%%%%%%%%%%%%%%%%|%

As discussed in the text, $\mu_{DC} \ge \Fphi$ to avoid tachyonic 
doubly-charged particles; however, given the form of 
\eq{Add.simple.superW.mass.DC}, it is evident the threshold associated with
the doubly-charged particles also preserves \abbr{AMSB}, which is true even if 
it is \emph{at} $\Fphi$.

%|%%%%%%%%%%%%%%%%%%%%%%%%%%%%%%%DO NOT EXCEED%%%%%%%%%%%%%%%%%%%%%%%%%%%%%%%|%

\begin{figure}
\includegraphics{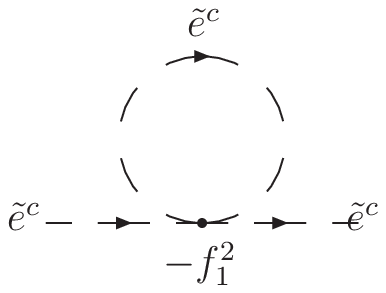}
\includegraphics{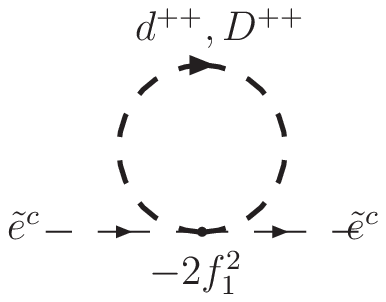}
\includegraphics{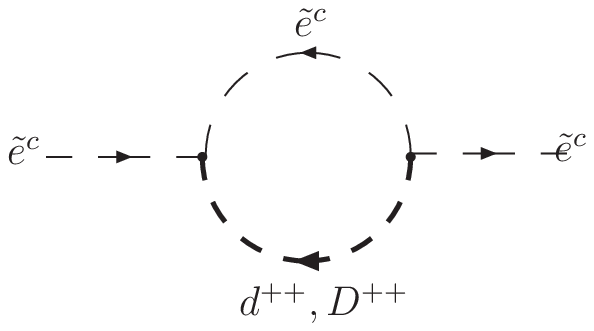}
\includegraphics{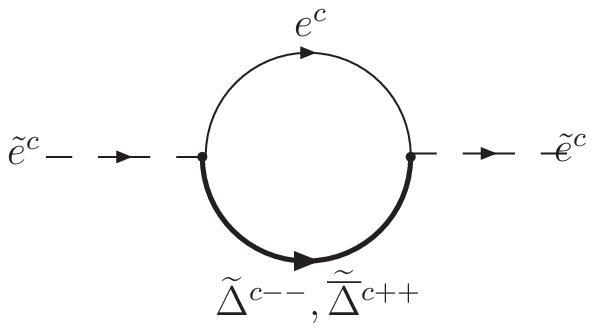}
\caption{One loop yukawa mediated contributions to the selectron from
integrating out the doubly-charged particles at $\mu_{DC}$.  The fields 
$d^{++}$, $D^{++}$ represent the mass eigenstates of the scalars $\DCmm$ and
$\DBarCpp$.}
\label{Fig:selectron.YMSB.delta.one.loop}
\end{figure}

But $\mu_{DC} \sim \Fphi$ has additional threshold corrections to the 
remaining low-scale particles that are important\cite{Dine:1996xk,Katz:1999uw}.
These effects are governed by the ratio
%
%|%%%%%%%%%%%%%%%%%%%%%%%%%%%%%%%DO NOT EXCEED%%%%%%%%%%%%%%%%%%%%%%%%%%%%%%%|%
%
\begin{equation}
\delta \equiv \frac{b_{DC}}{\mu_{DC}^2} = \frac{F_\phi}{\mu_{DC}}
\end{equation}
%
%|%%%%%%%%%%%%%%%%%%%%%%%%%%%%%%%DO NOT EXCEED%%%%%%%%%%%%%%%%%%%%%%%%%%%%%%%|%
%
which measures the splitting of the messenger scalar fields' masses due to 
\abbr{SUSY} breaking\footnote{If the scalar mass matrix of \eq{DC.Mass} has the 
eigenvalues $m_\pm^2$, then $(m_+^2 - m_-^2)/\mu_{DC}^2 = 2 \delta$}.  The
usual \abbr{AMSB} expressions for the low-scale particles are zero order in 
$\delta$, and are dominant for $\mu_{DC} \gg \Fphi$; however, for
$\mu_{DC} \sim \Fphi$ the one-loop yukawa-mediated contributions also
become important.  For the selectron, all such diagrams are shown in
\fig{selectron.YMSB.delta.one.loop}.
%
%|%%%%%%%%%%%%%%%%%%%%%%%%%%%%%%%DO NOT EXCEED%%%%%%%%%%%%%%%%%%%%%%%%%%%%%%%|%
%
%
%|%%%%%%%%%%%%%%%%%%%%%%%%%%%%%%%DO NOT EXCEED%%%%%%%%%%%%%%%%%%%%%%%%%%%%%%%|%
%
The sum of the graphs in \fig{selectron.YMSB.delta.one.loop} yield a scalar
mass-squared correction of
%
%|%%%%%%%%%%%%%%%%%%%%%%%%%%%%%%%DO NOT EXCEED%%%%%%%%%%%%%%%%%%%%%%%%%%%%%%%|%
%
\begin{equation}
\Delta m_{e^c}^2
	=	- \frac{2}{3} \frac{f_1^2 \mu_{DC}^2}{16 \pi^2} \delta^4 
	\sim	- \frac{2}{3} \frac{\Fphi^2}{16 \pi^2} f_1^2,
\label{Eq:YMSB.Slepton.Masses}
\end{equation}
%
%|%%%%%%%%%%%%%%%%%%%%%%%%%%%%%%%DO NOT EXCEED%%%%%%%%%%%%%%%%%%%%%%%%%%%%%%%|%
%
where the second expression takes $\mu_{DC}$ around $\Fphi$.  This 
expression is always negative and larger in magnitude than the \abbr{AMSB} 
expressions, which are suppressed by an additional factor of $1/16\pi^2$.

%|%%%%%%%%%%%%%%%%%%%%%%%%%%%%%%%DO NOT EXCEED%%%%%%%%%%%%%%%%%%%%%%%%%%%%%%%|%

At this stage it would appear that combining the seesaw mechanism with 
\abbr{AMSB} has actually made the problem worse, since the sleptons are now
`more negative' by a factor of $16\pi^2$.  This is not, however, the situation
because the model itself permits additional terms that are not expressed in
the superpotential.  In fact, the full model of 
Eqs.~\eqn{SuperW.SUSYLR}--\eqn{SuperW.SUSYLR.nr.gsv} allow the k\"ahler 
potential term
%
%|%%%%%%%%%%%%%%%%%%%%%%%%%%%%%%%DO NOT EXCEED%%%%%%%%%%%%%%%%%%%%%%%%%%%%%%%|%
%
\begin{equation}
\mathcal{K} 
	\supset	k \frac{\phi^\dagger}{\phi} 
			\Tr\inp{\Delta\DeltaBar + \DeltaC\DeltaBarC}
\label{Eq:Add.kahler.bilinear}
\end{equation}
%
%|%%%%%%%%%%%%%%%%%%%%%%%%%%%%%%%DO NOT EXCEED%%%%%%%%%%%%%%%%%%%%%%%%%%%%%%%|%
%
with $k$ an order one constant.

A term such as \eq{Add.kahler.bilinear} has been studied 
before\cite{Nelson:2002sa,Hsieh:2006ig}, and it was pointed out that it yields
an effective superpotential term of
%
%|%%%%%%%%%%%%%%%%%%%%%%%%%%%%%%%DO NOT EXCEED%%%%%%%%%%%%%%%%%%%%%%%%%%%%%%%|%
%
\begin{equation}
\intOp[4]{\theta} \mathcal{K}
	\supset \intOp[4]{\theta} k \frac{\phi^\dagger}{\phi} 
						\Tr\inp{\DeltaC\DeltaBarC}
	=	\intOp[2]{\theta} k \frac{\Fphi^\dagger}{\phi} 
						\Tr\inp{\DeltaC\DeltaBarC}
\;\;\leftrightarrow\;\;
\mathcal{W}
	\supset k \frac{\Fphi^\dagger}{\phi} \Tr\inp{\DeltaC\DeltaBarC}.
\label{Eq:Add.effective.superW.mass.deltaC}
\end{equation}
%
%|%%%%%%%%%%%%%%%%%%%%%%%%%%%%%%%DO NOT EXCEED%%%%%%%%%%%%%%%%%%%%%%%%%%%%%%%|%
%
The presence of this effective \abbr{SUSY} mass term then alters the mass
matrix for the doubly-charged particles given in \eq{DC.Mass} to
%
%|%%%%%%%%%%%%%%%%%%%%%%%%%%%%%%%DO NOT EXCEED%%%%%%%%%%%%%%%%%%%%%%%%%%%%%%%|%
%
\begin{equation}
\mathcal{M}_{DC} =
\begin{pmatrix}
\bigl|\mu_{DC} + k \Fphi^\dagger\bigr|^2
				& \mu_{DC} \Fphi - \abs{k \Fphi}^2	\\
\mu_{DC}^\dagger \Fphi^\dagger - \abs{k \Fphi}^2
				& \bigl|\mu_{DC} + k \Fphi^\dagger\bigr|^2
\end{pmatrix}
\label{Eq:Add.DC.mass.matrix}
\end{equation}
%
%|%%%%%%%%%%%%%%%%%%%%%%%%%%%%%%%DO NOT EXCEED%%%%%%%%%%%%%%%%%%%%%%%%%%%%%%%|%
%
with $\mu_{DC} \sim v_R^2/\MNP$ as before.  Since $k$ and $\mu_{DC}$ are free
parameters ($k$ is an arbitrary $\mathcal{O}(1)$ constant while $\mu_{DC}$ 
depends on the non-renormalizable couplings), \eq{Add.DC.mass.matrix} may be 
tuned so that all the fields are at $\MSUSY$:
%
%|%%%%%%%%%%%%%%%%%%%%%%%%%%%%%%%DO NOT EXCEED%%%%%%%%%%%%%%%%%%%%%%%%%%%%%%%|%
%
\begin{equation}
\begin{aligned}[b]
\bigl|\mu_{DC} + k \Fphi^\dagger\bigr|^2	
	& \sim \frac{\abs{\Fphi}}{16 \pi^2}
	\\
\mu_{DC} \Fphi - \abs{k \Fphi}^2	
	& \sim \pfrac{\abs{\Fphi}}{16 \pi^2}^2
\end{aligned}
\label{Eq:Add.DC.tuning}
\end{equation}
%
%|%%%%%%%%%%%%%%%%%%%%%%%%%%%%%%%DO NOT EXCEED%%%%%%%%%%%%%%%%%%%%%%%%%%%%%%%|%
%
The tunings \eq{Add.DC.tuning} permit both the doubly-charged fermions and the
doubly-charged scalars to remain in the theory to the TeV scale and retain the
\abbr{AMSB} trajectory for all the particles.  A similar argument allows the 
left-handed triplets to persist until $\MSUSY$. While both the doubly-charged 
scalars and fermions survive to the TeV scale, the muonium-antimuonium 
constraints given in \eq{DC.muonium.constraint} still force these particles' 
masses to be at or above $2$ TeV.  If they reside right near this lower bound, 
the \abbr{LHC} may produce both doubly-charged scalars and fermions (as opposed
to just the scalars as presented in the paper).

%|%%%%%%%%%%%%%%%%%%%%%%%%%%%%%%%DO NOT EXCEED%%%%%%%%%%%%%%%%%%%%%%%%%%%%%%%|%

Because this new particle content survives to the TeV scale, the \abbr{AMSB}
expression may be utilized at that scale to determine the soft masses.  The
presence of the new yukawa couplings $f$ and $f_c$ for the sleptons will
then cause them to be positive.  In the analysis of \Sec{Sparticle.Masses}, 
these \abbr{AMSB} expressions were evaluated at $\Fphi$ for both squarks and 
sleptons, then used as boundary conditions to evolve the masses down to 
$\MSUSY$.  As the parameters do not run significantly from $\Fphi$ to 
$\MSUSY$ (it is only two orders of magnitude), the numerical results presented
in the paper remain valid within the expected uncertainty.

%|%%%%%%%%%%%%%%%%%%%%%%%%%%%%%%%DO NOT EXCEED%%%%%%%%%%%%%%%%%%%%%%%%%%%%%%%|%

\section*{Note} After this paper was published, the authors were informed of 
\cite{deAlwis:2008aq} which discusses an alternative scenario to avoiding 
tachyonic sleptons.  The authors regret this omission and their oversight which
prevented it appearing in the printed paper.

%%%%%%%%%%%%%%%%%%%%%%%%%%%%%%%%%%%%%%%%%%%%%%%%%%%%%%%%%%%%%%%%%%%%%%%%%%%%%%%
%[[[[[[[[[[[[[[[[[[[[[[[[[[[[[[[[[ REFERENCES ]]]]]]]]]]]]]]]]]]]]]]]]]]]]]]]]]
%%%%%%%%%%%%%%%%%%%%%%%%%%%%%%%%%%%%%%%%%%%%%%%%%%%%%%%%%%%%%%%%%%%%%%%%%%%%%%%

%|%%%%%%%%%%%%%%%%%%%%%%%%%%%%%%%DO NOT EXCEED%%%%%%%%%%%%%%%%%%%%%%%%%%%%%%%|%
\bibliography{%
\bibpath/amsb,%
\bibpath/susy,%
\bibpath/susy-dark_matter,%
\bibpath/susylr,%
\bibpath/susy-extended_higgs,%
\bibpath/nmssm,%
\bibpath/susy-ext_to_superspace,%
\bibpath/flavor_violation,%
\bibpath/computer_codes,%
\bibpath/gmsb,%
\bibpath/neutrinos,%
\bibpath/vacuum-color_violation,%
\bibpath/colliders,%
\bibpath/extradimensions%
}

%|%%%%%%%%%%%%%%%%%%%%%%%%%%%%%%%DO NOT EXCEED%%%%%%%%%%%%%%%%%%%%%%%%%%%%%%%|%

%%%%%%%%%%%%%%%%%%%%%%%%%%%%%%%%%%%%%%%%%%%%%%%%%%%%%%%%%%%%%%%%%%%%%%%%%%%%%%%
%[[[[[[[[[[[[[[[[[[[[[[[[[[[[[[[ END REFERENCES ]]]]]]]]]]]]]]]]]]]]]]]]]]]]]]]
%%%%%%%%%%%%%%%%%%%%%%%%%%%%%%%%%%%%%%%%%%%%%%%%%%%%%%%%%%%%%%%%%%%%%%%%%%%%%%%
\end{document}